\documentclass[traditabstract]{aa}

\usepackage{natbib}
\bibpunct{(}{)}{;}{a}{}{,} 

\usepackage{graphicx}
\usepackage{txfonts}
\usepackage{ulem}
\usepackage[breaklinks,colorlinks=true,citecolor=black,linkcolor=black,urlcolor=blue,bookmarks=true]{hyperref}
\newcommand{\kms}{km\,s$^{-1}$}

\newcommand{\FeH}{\left[\mathrm{Fe}/\mathrm{H}\right]}
\newcommand{\HI}{\ion{H}{i}}

\newcommand{\catRec}{6\,640}
\newcommand{\catObj}{2\,335}
\newcommand{\catArt}{430}

\begin{document}

\title{HyperLEDA. III. The catalogue of extragalactic distances}
\titlerunning{HyperLEDA: distance catalogue}

\author{
Dmitry Makarov\inst{1,2}\fnmsep\thanks{\email{dim@sao.ru}}
\and
Philippe Prugniel\inst{1} 
\and
Nataliya Terekhova\inst{3}
\and
H\'el\`ene Courtois\inst{4} 
\and
Isabelle Vauglin\inst{1}
}
\authorrunning{Makarov et al.}

\institute{
Universit\'e Lyon~1, Villeurbanne, F-69622, France; CRAL, Observatoire de Lyon, St. Genis Laval, F-69561, France; CNRS UMR 5574, France
\and
Special Astrophysical Observatory, Nizhniy Arkhyz, Karachai-Cherkessia 369167, Russia
\and
Sternberg Astronomical Institute, Moscow State University, Universitetsky pr., 13, Moscow 119991, Russia
\and
Universit\'e Claude Bernard Lyon I, Institut de Physique Nucleaire, F-69622 Lyon, France
}

\date{Received date / accepted date}

\abstract{
We present the compilation catalogue of redshift-independent distances included in the HyperLEDA database. 
It is actively maintained to be up-to-date, and the current version counts \catRec{} distance measurements for \catObj{} galaxies compiled from \catArt{} published articles.
Each individual series is recalibrated onto a common distance scale based on a carefully selected set of high-quality measurements.
This information together with data on \HI{} line width, central velocity dispersion, magnitudes, diameters,
and redshift is used to derive a homogeneous distance estimate and physical properties of galaxies, 
such as their absolute magnitudes and intrinsic size.
}


\maketitle

\section{Introduction}

Measuring distances is a long-standing and on-going goal of astronomy.
Significant milestones were the measurement of the distance to the Moon by \citet[discussed by \citealp{Toomer1974}]{Hipparchus}, 
and, in the modern era, the determination of the Cepheid distances, which unveiled the true nature of galaxies \citep{Hubble1926}.

Although the discovery of a relation between the distance and the radial velocity of galaxies \citep{Hubble1929} 
built the paradigm of an expanding Universe and gave a simple `proxy' to the evaluation of extragalactic distances,
redshift-independent distances are still vital.
In particular, the deviation of radial velocities from the Hubble law gives us information about cosmic flows \citep{TSK2008}
and mass distributions \citep{KKMT2009,CHTG2012}.
The distances are also obviously needed to fix the Hubble constant, $H_0$, and they are crucial to constrain the cosmological parameters.
Most notably, the use of distant supernovae as standard candles led to the discovery of the accelerated expansion of the Universe \citep{Riess+98,Perlmutter+99}.
These observations together with other evidence form the modern standard cosmological model, according to which our Universe is mostly `dark'.
It consists of about 73\,\% of dark energy, 22\,\% dark matter, and only $\sim$5\,\% of baryon matter \citep{WMAP7yr}.

There are many indicators from which extragalactic distances can be derived.
The best-calibrated and most precise methods are usually observationally expensive and limited to the nearby Universe, 
for example the period-luminosity relation for Cepheids.
At larger distance, most indicators are based on scaling laws of galaxies, for example, the well-known Tully-Fisher relation.
Because no single technique can work on all scales, a consistent system is constructed step-by-step, from nearby to distant objects. 
In this process, each distance indicator is calibrated with respect to those available at nearest scales.
For this purpose it is necessary to maintain a database of precise measurements that can be homogenized into a common scale that can be used as standards.

The goal of the present work is to describe the galaxy distance catalogue maintained within the HyperLEDA database\footnote{\url{http://leda.univ-lyon1.fr/}}.
It compiles distances published in the astronomical literature and provides precise descriptions of these measurements.
In addition to this catalogue, HyperLEDA contains a consistent body of data from the literature on photometry, 
\HI{} line width, internal stellar kinematics, and other characteristics.
All these catalogues are combined and are corrected for systematic effects to provide a homogenized description of galaxies.
In particular, the distances to galaxies are derived using the present catalogue supplemented by distance indicators 
based on other data (Tully-Fisher, Faber-Jackson, and Fundamental Plane relations).

HyperLEDA, including the distance catalogue, is also used for other projects.
One of them is the Catalogue of the Local Volume galaxies\footnote{\url{http://www.sao.ru/lv/lvgdb/}} \citep[hereafter LVG]{CNG,UNGC},
which gives distances, velocities, and physical characteristics for galaxies within 11\,Mpc (at present 869 galaxies).
The LVG results from a detailed and careful analysis of the data collected in databases and in the literature.
Special attention is given to cleaning the data from artefacts, Galactic objects, and doubtful measurements.
This makes LVG much more than a mere subsample extracted from general databases.
Ongoing efforts make LVG the most complete sample of nearby galaxies, that is virtually free of contamination.

Another project using the present catalogue is the Extragalactic Distance Database\footnote{\url{http://edd.ifa.hawaii.edu/}} \citep[EDD]{EDD}.
It is intended to collect information related to the distance determination within 100--200\,Mpc.
EDD combines original observations with published data and compilations.
Its Tully-Fisher relation is derived from a uniform analysis of the \HI{} line width \citep{EDD:HI} and 
photometry \citep{CosmicFlows:HawaiiPhoto},
whereas the zero-point calibration is derived from original distance determination 
using the colour-magnitude diagrams of nearby galaxies from the Hubble Space Telescope \citep{EDD:CMD}.
The calibration sample for short distances is supplemented by the present catalogue.

The NASA/IPAC Extragalactic Database\footnote{\url{http://ned.ipac.caltech.edu/}} (NED) 
maintains an on-line compilation of redshift-independent extragalactic distances from the literature (NED-D).
It collects huge sets of published measurements based on both primary methods such as Cepheids or Type Ia supernova (SN\,Ia), 
and secondary indicators such as Tully-Fisher or Fundamental Plane relations.

The present database is more focused on the nearby galaxies to provide the best standards to calibrate distance indicators.
In Section~\ref{sect:hl} we briefly describe the organization and content of HyperLEDA,
Sect.~\ref{sect:catalogue} describes the distance catalogue, 
Sect.~\ref{sect:homogenization} focuses on the homogenization of different distance determination methods,
and Sect.~\ref{sect:conclusion} draws conclusions.

\section{HyperLEDA database}
\label{sect:hl}

HyperLEDA \citep{HyperLEDA1,HyperLEDA2} takes its roots in the Lyon-Meudon Extragalactic database \citep[LEDA]{LEDA}, 
which was created in 1983,
and in Hypercat, which started as the observational catalogue on kinematics of early-type galaxies \citep{PS1996}.
These databases were joined in 2000.
Historically, the LEDA database was used for preparation of the Third Reference Catalogue of Bright Galaxies \citep[RC3]{RC3}.
Now HyperLEDA is operated by the Observatoire de Lyon (France) and by the Special Astrophysical Observatory (Russia).

The basis of HyperLEDA is a set of compilation catalogues that are maintained on a regular basis. 
Each of them collects specific data on astronomical objects from the literature.
The maintenance of these catalogues involves efforts to provide accurate cross-identifications of the celestial sources, 
a clear description of each series of measurements (including a documentation on the precision and systematic errors), 
and a flagging of doubtful or erroneous data.

The main part, the so-called LEDA catalogue, combines information from the compilations to provide a uniform and self-consistent description of all objects.
It consists of homogenized observations such as the total apparent magnitude in the $B$ band, apparent diameter, and redshift, 
as well as physical parameters including the absolute magnitude and the maximum rotation velocity.
The homogenization process is based on the description of each series of measurements (for example how to apply an aperture correction)
and on a statistical comparison of these series over the whole catalogue (for example how to correct zero-points).
The physical values are determined from the homogenized apparent data, taking into account different kinds of corrections
(Galactic extinction, object inclination, line-width broadening because of redshift, and others).
The homogenization and parameter determination are described by \citet{LEDA7}.
More recently, some aspects where revised by \citet{HyperLEDA1,HyperLEDA2}. 
The online documentation provides updated details.

The principal compilation catalogues of HyperLEDA are described below.

\begin{description}
\item[Astrometry and Designation:] 
  These catalogues contain 8\,177\,892 celestial positions and 6\,878\,482 designations for 3\,730\,169 objects.
  The designations stored in the database comply with the IAU recommendation\footnote{\url{http://cdsweb.u-strasbg.fr/Dic/iau-spec.html}}.
  These catalogues are fundamental for the unique identification of objects.
  They are described by \citet{HyperLEDA1}.

\item[Gas kinematics:]
  This compilation of \HI{} data associated with optical counterparts is the origin of the LEDA database \citep{HyperLEDA2}.
  At present, the catalogue gathers 113\,086 measurements of a \HI{} line width or a maximum rotation curve for 37\,377 galaxies.
  The catalogue is used to derive the homogenized physical maximum rotation velocity, \textsc{vrot}, corrected for inclination.
  Together with photometric information it is used to estimate distance
  using the Tully-Fisher relation \citep{TullyFisherRelation} for spiral galaxies.

\item[Group membership:] 
  HyperLEDA also indexes multiple systems, including pairs, triplets, groups, and clusters. 
  The catalogue collects information about groups and members of groups from the literature.
  The user interface allows us to determine the groups that an object is member of and, reciprocally, to find the members of a particular group. 
  The database contains 19\,829 groups, and the compilation gives 391\,247 membership records for 370\,999 objects.
  In addition and independently of this, the web interface allows us to obtain a list of objects grouped to a specific target
  on the basis of proximity in projection on the sky and in redshift space, using the algorithm described by \citet{PGM1999}.

\item[Mg$_2$ line strength indexes:] 
  The catalogue of published absorption-line Mg$_2$ indices of galaxies and globular clusters \citep{GP1998} 
  currently contains 9\,883 measurements for 3\,271 objects.

\item[Morphology:] 
  The catalogue collects the RC3 numerical types, t, in de Vaucouleur's scale \citep[][see section\,3.3]{RC3},
  which are combined with information on presence of bar, ring, multiplicity, 
  or interaction to build the standard morphological classification.
  At the moment, HyperLEDA compiles 232\,305 morphological codes for 112\,572 objects.
  Most of these classifications result from a visual inspection of optical images.
  Recently an effort was made to calibrate an automatic classification algorithm and
  give a homogeneous classification for 4\,458 bright galaxies from the Sloan Digital Sky Survey \citep{EFIGI}.

\item[Nature:] 
  Although HyperLEDA is intended to be a database of extragalactic objects, it is not possible to restrict it to galaxies alone.
  Often, the physical nature of an object is not known at the time of discovery,
  and it formally only appears as \textit{an extended source}, 
  while its extragalactic nature is merely a presumption. 
  Other observations may confirm or change the earlier classification.
  Because the data are never excluded from the database, 
  HyperLEDA contains virtually all kinds of celestial objects. 
  The catalogue gives 5\,014\,711 nature classifications provided in the literature or made by the HyperLEDA team. 
  The homogenization process combines the different classifications for each object, 
  automatically chooses the most precise or assigns an {\it undetermined nature} in case of inconsistency. 
  These cases are marked for human control to possibly solve the inconsistency. 
  HyperLEDA identifies 3\,358\,587 galaxies and 372\,941 stars and objects of other nature.

\item[Nuclear activity classification:] 
  This catalogue compiles information about the detection of signatures of activity in the centre of galaxies 
  (active galactic nuclei, AGN,  or star formation) for 88\,421 objects \citep{HL:AGN}.
  In particular, it contains the V\'{e}ron catalogue of quasars and 
  AGN\footnote{\url{http://heasarc.gsfc.nasa.gov/W3Browse/all/veroncat.html}} \citep{VeronCat13}.

\item[Photometry:] 
  The catalogue presents the compilation of 28\,208\,556 apparent fluxes for 3\,605\,940 objects from the literature and from surveys.
  Mostly they are optical, near infra-red, and \HI{} magnitudes.
  These values are stored as is, without correction for Galactic extinction or any other effects.
  This catalogue is supported with some aperture photometry (useful for calibrating observations of large nearby galaxies), 
  photometric profiles, fitted growth curves, and colours \citep{PH1998}.
  The surface brightness is also given for 27\,761 galaxies.
  \citet{LEDA5,LEDA7} described the reduction of apparent $B$ magnitude to the RC3-system, $B_\mathrm{T}$.

\item[Redshift:]
  The catalogue collects 2\,387\,020 redshifts in the heliocentric rest frame for 1\,296\,804 objects.
  The data are presented in units of \kms{} as $cz$, where $z$ is the redshift and $c=299\,792.458$ \kms{} is the speed of light.
  For low redshift, $z \ll 1$, the value $cz$ can be treated as the radial velocity of the object in the so-called optical convention.
  The homogenization of redshifts, transformation to different rest frames such as to the cosmic microwave background dipole or centroid of the Local Group 
  as well as correction for Virgo infall is given by \citet{LEDA7}.

\item[Sizes and position angles:] 
  These catalogues compile information on major and minor diameters of objects as well as the position angle of the major axis counted from north to east.
  The data are gathered for different pass-bands, various isophote levels, and different measurement techniques.
  The catalogue collects 11\,016\,356 diameter measurements for 3\,009\,469 objects.
  The position angle is known for 2\,763\,881 objects.
  The reduction to the standard system, corresponding to the size at the 25 $B$-mag arcsec$^{-2}$ isophote, is given by \citet{LEDA:Size,LEDA7}.

\item[Spatially resolved kinematics of galaxies:] 
  The bibliographical catalogue assembles information on 15\,197 publications 
  on rotation curve observations for 3\,860 galaxies \citep{PZBS1998}.

\item[Stellar kinematics:] 
  The catalogue, presented by \citet{PS1996}, contains 23\,797 observations 
  of a central velocity dispersion for 16\,927 objects and 
  1\,668 measurements of maximum rotation of stellar populations for 832 early-type galaxies. 
  It is used in HyperLEDA to derive distances through the Faber-Jackson and the Fundamental Plane relations,
  after homogenization and combination with photometric data.

\end{description}

\noindent
This article describes the catalogue of redshift-independent distance estimates.

\begin{figure}
\centerline{
\includegraphics[width=0.45\textwidth,clip]{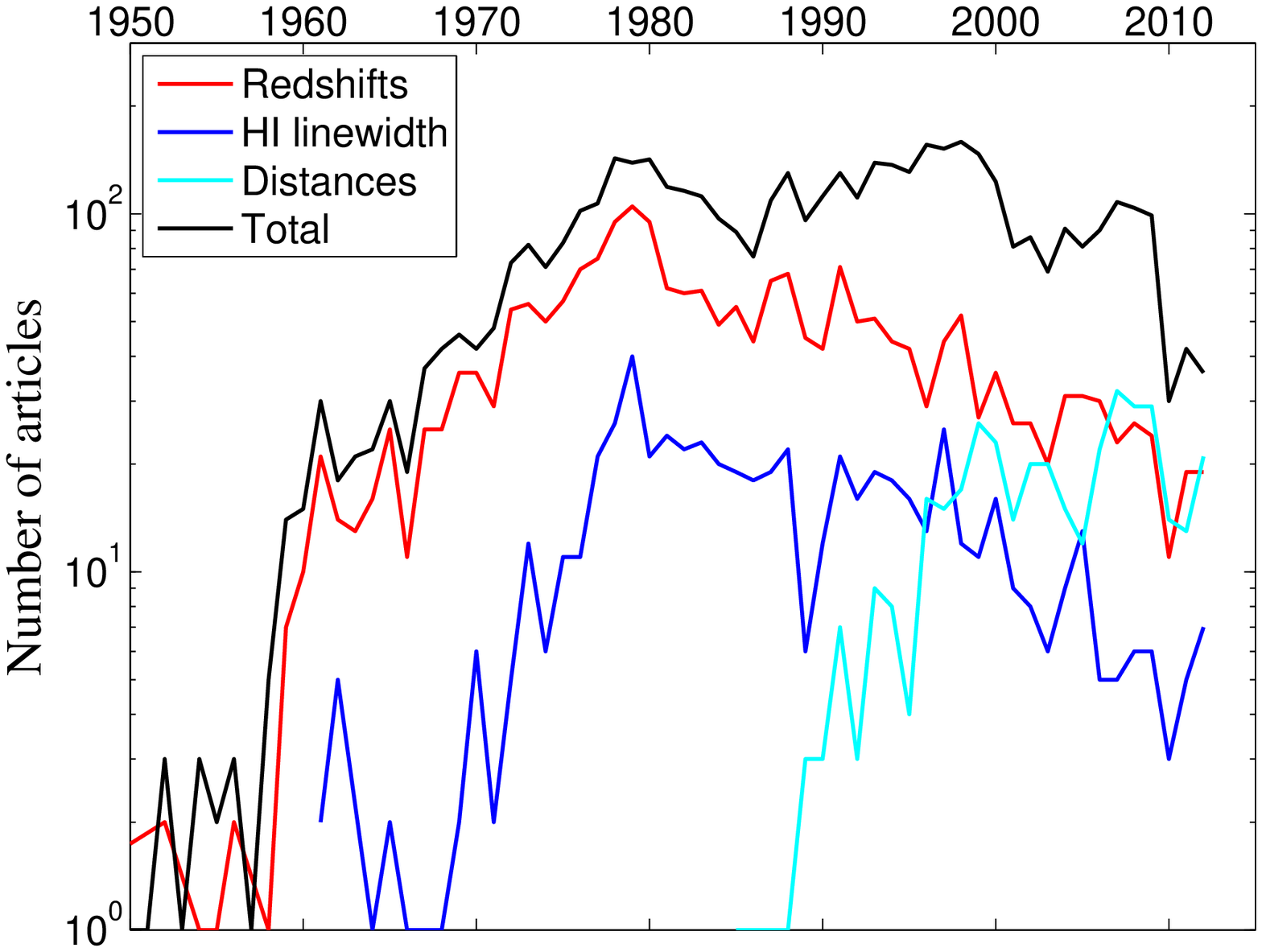}
}
\centerline{
\includegraphics[width=0.45\textwidth,clip]{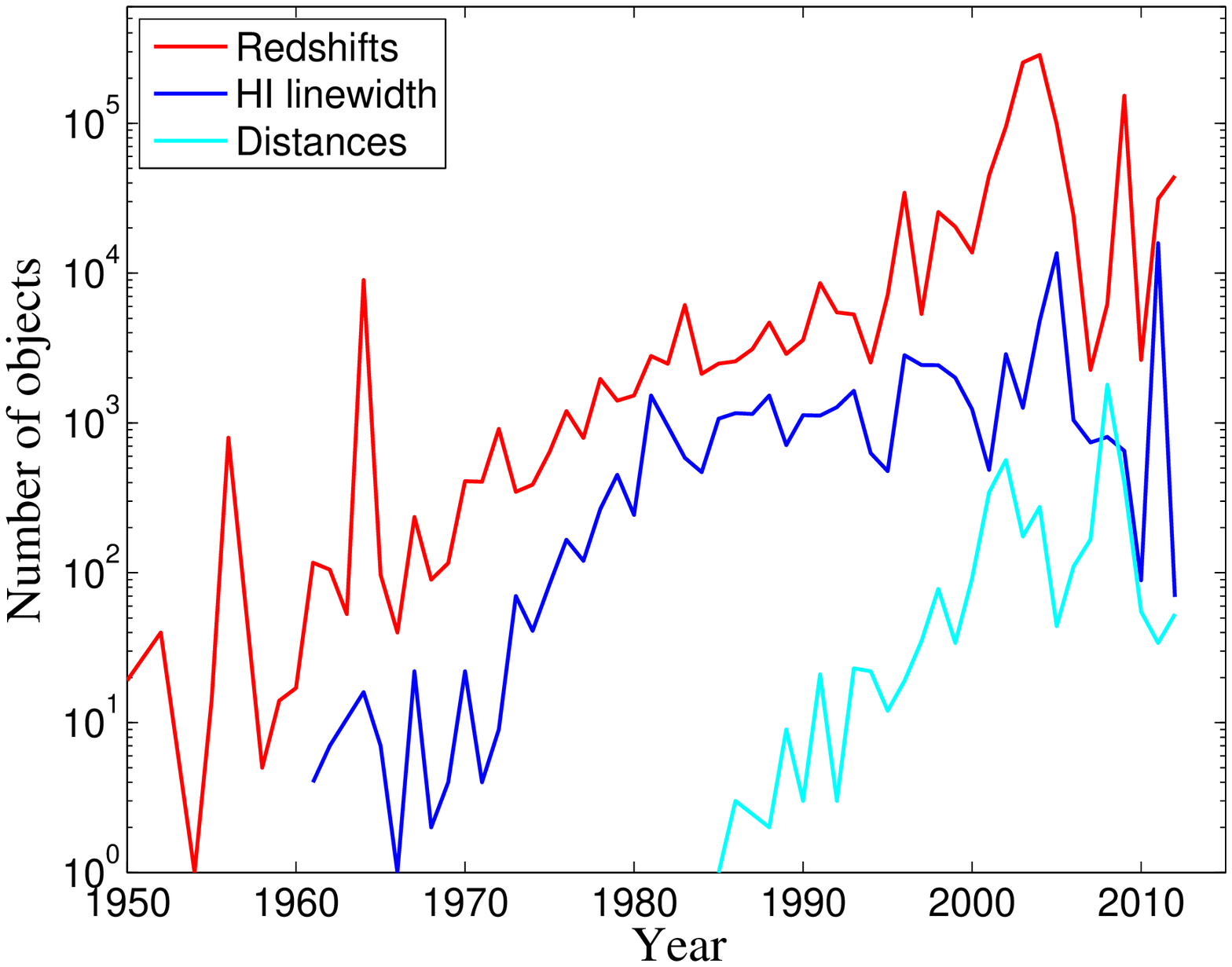}
}
\caption{
Year-by-year statistics of data ingestion and processing in HyperLEDA.
The top panel shows the number of processed articles per year.
The black line corresponds to the whole database, 
while individual compilation catalogues are shown by different colours:
the redshift catalogue is red, the \HI{} line width is blue, and the distance catalogue is cyan.
The bottom panel shows the number of individual objects processed in the given year.
}
\label{f:year}
\end{figure}

\begin{figure*}
\centerline{
\includegraphics[width=0.95\textwidth,clip]{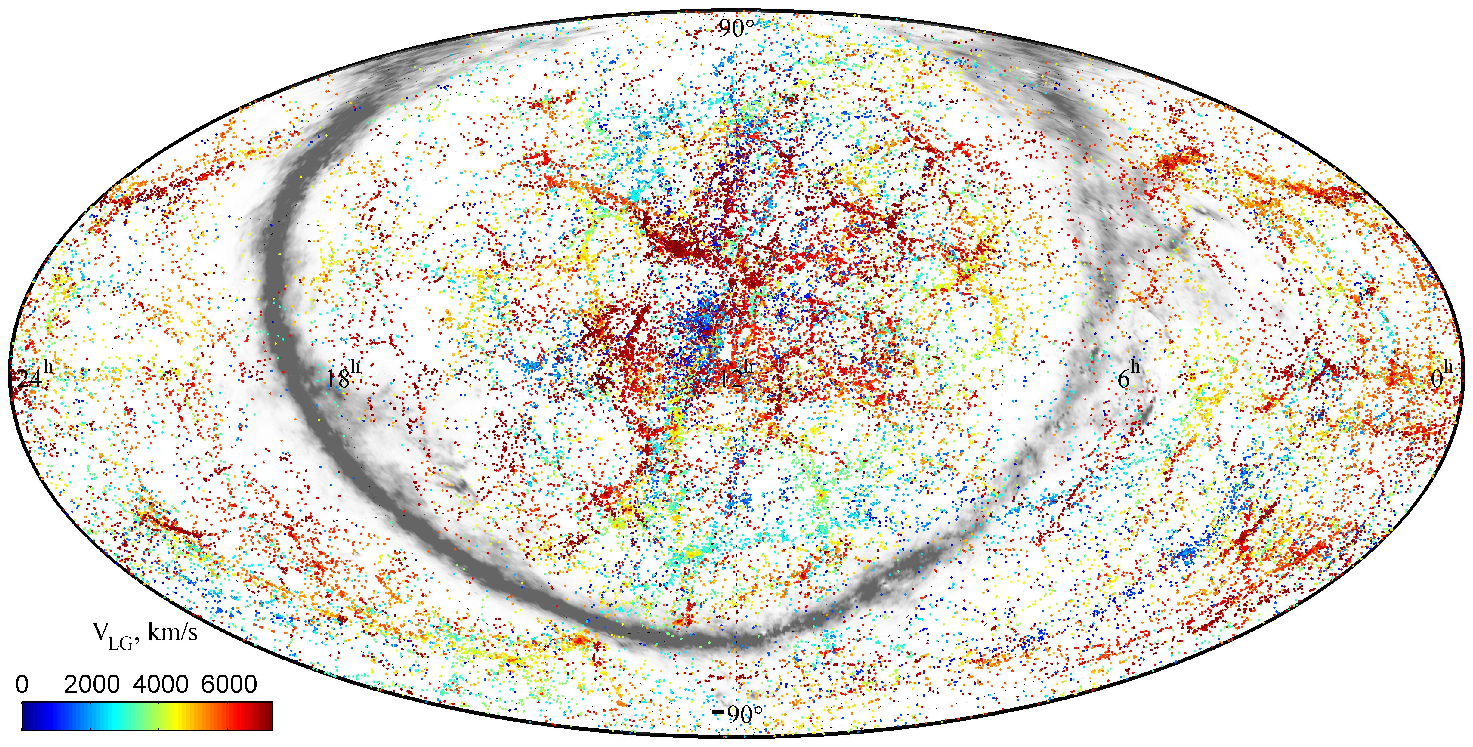}
}
\centerline{
\includegraphics[width=0.95\textwidth,clip]{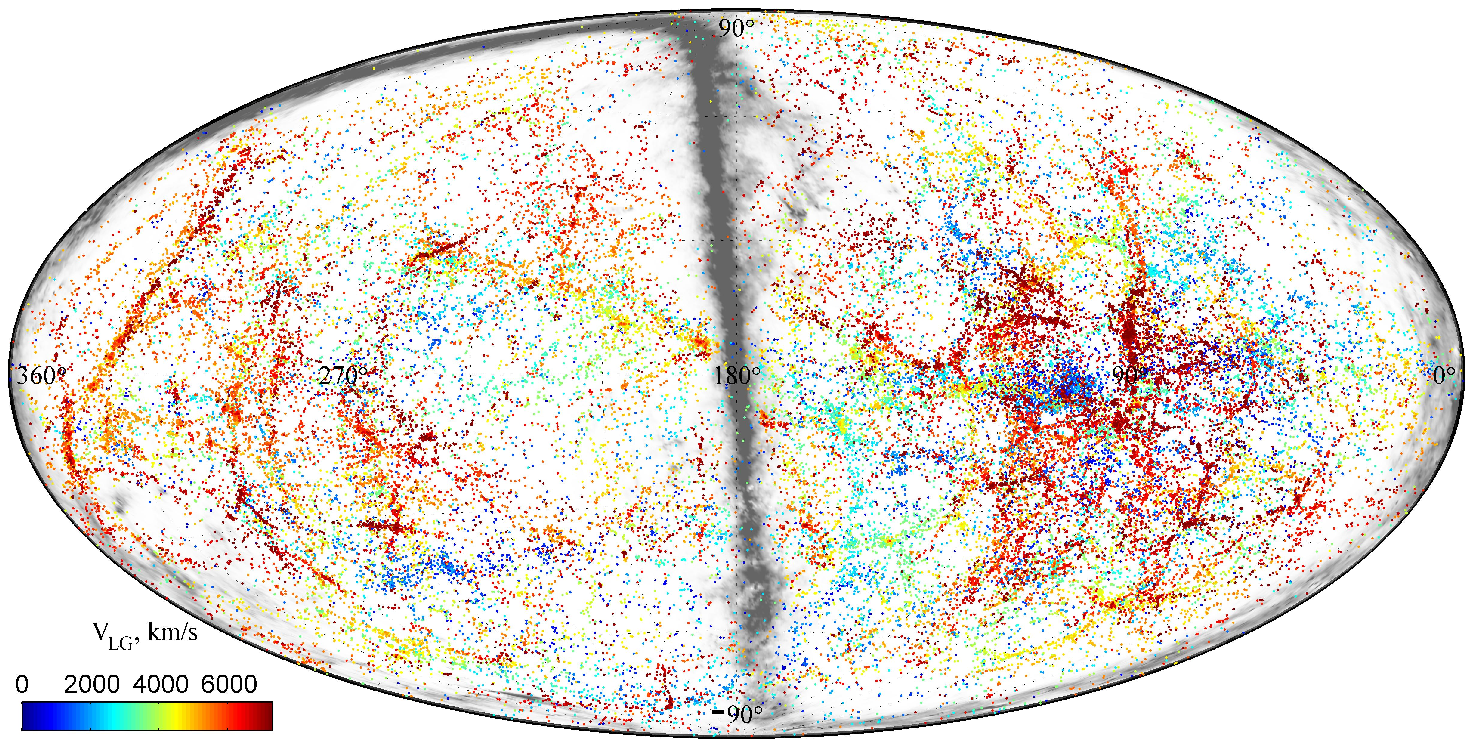}
}
\caption{
All-sky distribution of about 60\,000 galaxies within 100\,Mpc from HyperLEDA in the equatorial (top panel)
and the supergalactic coordinates (bottom panel).
The galaxy redshifts are colour-coded from blue for nearby objects to brown for distant ones.
The Zone of Avoidance in the Milky Way is shown by the grey clumpy clouds. 
The darkest regions correspond to the highest absorption, according to the \citet{Schlegel+1998} extinction map.
}
\label{f:sky100}
\end{figure*}

The public mirrors of HyperLEDA are operated with the \textsc{pleinpot} software.
Both the software and data are available as a free package\footnote{\url{http://leda.univ-lyon1.fr/install/mirror.html}} 
for a Unix/Linux system, which enables installing a private mirror of the system.
The web interface gives access to a number of tools. 
The principal ones are the searches by designation, near a position (also called cone searches), and by parameters 
using Structure Query Language (SQL)-like commands.

The search by designation prompts the user for the name of an object or a list of names. 
These names are searched for in the designation catalogue, taking into account a range of variants in their spelling
(for example abbreviations or alternatives to the catalogue acronym) and accepting some wildcard characters. 
In addition, a specific syntax allows searching for the nearest object to a given position on the sky. 
The search can be applied to the LEDA catalogue or to any of the other compilations.

The cone search takes a position on the sky and a radius to limit the surrounding area.  
A variety of images of the field can be displayed with a possible overlay that the database objects use, 
for instance, the Aladin\footnote{\url{http://aladin.u-strasbg.fr/}} applet \citep{Aladin} from the Centre de Donn\'ees de Strasbourg.

The SQL search gives a flexible access to the LEDA catalogue. 
The user can define her or his own SQL constraints and retrieve any desired fields or valid SQL expressions involving fields.

Some other search modes and tools are also available from the interface, but are not described in this short introduction.

Figure~\ref{f:year} illustrates the year by year statistics on data ingestion and processing in HyperLEDA.
The top panel shows the number of surveyed articles sorted by on the year of publication for the whole database
and for some compilations, namely the redshift, the gas kinematics, and the distance, which last is the object of this article.
The bottom panel shows the number of added objects in one year.

Figure~\ref{f:sky100} illustrates the distribution on the sky of the galaxies with known redshift up to $V_\mathrm{LG} < 7\,300$ \kms{},
which corresponds to a distance of about 100\,Mpc.
$V_\mathrm{LG}$ denotes the radial velocity with respect to the Local Group centroid, as defined in the HyperLEDA online documentation.
The colour of the symbols indicates the redshift.
The map exhibits the well-known filamentary structure of galaxy distribution, 
which connects the massive clusters and depicts the low-density regions.
A detailed description of structures in a similar volume of the local Universe was recently carried out by \citet{Cosmography}.
Despite the modern surveys, data in the Zone of Avoidance at low Galactic latitudes are highly incomplete.
Galaxy extinction is shown as a grey clumpy belt in Fig.~\ref{f:sky100}.
The footprints of the different galaxy surveys are barely detectable, 
which confirms that in the nearby Universe the coverage of the sky by different redshift surveys is reasonably uniform 
(i.e.\ the completeness is about the same in any direction, except for the Galactic plane).
In contrast, farther away than 100\,Mpc, the sky coverage is not uniform because some regions are surveyed to a deeper level than others. 
In particular, the catalogue is much more complete in the region surveyed by the Sloan Digital Sky Survey \citep{SDSSdr10} than in the rest of the sky.
Despite this, the general structures such as walls, filaments, clusters, and voids are still visible, 
but statistical uses of the database should take into account these various selection biases.

\section{Description of the distance catalogue}
\label{sect:catalogue}

\subsection{Organisation of the compilation}

The distance catalogue in HyperLEDA collects the redshift-independent distance measurements published in the literature.
It can be either original distance measurements, re-calibrations, or even compilations of data.
Since it is difficult with limited resources to maintain an up-to-date complete literature survey, 
we are prioritizing the high-quality measurements that are potentially the most interesting standards for calibrating other indicators. 
The inclusion of lower priority datasets may be delayed by a few years.

An important step in adding a new source of data to the catalogue is to cross-identify its objects with HyperLEDA. 
An object in a publication is usually identified by one or several names and by coordinates (when the IAU recommendations are followed).
In most cases it allows an automatic identification when the position and designations consistently match a unique object in HyperLEDA. 
In the opposite case, when there is no correspondence, a new object is added into the database.
However, when the information is inconsistent (e.g.\ the position does not correspond to some of the designations), 
or when several HyperLEDA objects match the specification, manual handling is required. 
Although these cases are quite rare, typically 0.5--1\,\% for the distance catalogue, 
the manual cross-identification consumes most of the human resource required for maintaining the database. 

The goal of the cross-identification is to associate a unique internal identification number 
(\textsc{pgc}, see below) with each object in the new article.
Then, to add the new data in HyperLEDA, we edit and execute \textit{ingestion rules} 
that describe how to convert the published data into the internal HyperLEDA storage.
This conversion involves very little alteration, for example, the transformation of a published linear distance to a stored distance modulus.

In addition to the cross-identification, the other most time-consuming task is to code a description of the measurement protocol. 
This meta-information is essential for the subsequent measurement homogenization. 
This information is stored in a specific table where each entry describes a calibration 
used for distance measurements (the calibration parameter table, see below).

\subsection{Structure of the catalogue}

The distance catalogue currently consists of two main blocks: the measurement and the calibration tables. 

The measurement table stores the actual published distance determinations, merging the data from the all sources in a single table. 
The fields of the table are chosen to keep the published values as completely as possible with minimal changes
and to maintain an easy traceability back to the original paper. 

The calibration table contains ancillary information describing the calibration reference (zero point) of a series of measurements.

In addition to these tables, we use the HyperLEDA bibliographical reference catalogue for publication linkage.
It is shared by all the HyperLEDA catalogues.
Each article in the bibliographic table as far as possible is associated with its standard code, commonly called \textsc{bibcode}. 
This allows one to establish a connection with other databases, 
in particular with the SAO/NASA Astrophysical Data System\footnote{\url{http://cdsads.u-strasbg.fr/}} (ADS), 
where the full text of the original publication is available.
The bibliographic table also contains some information about the history of the addition of the reference to HyperLEDA.

Finally, meta-data tables provide the description of the catalogue itself 
as well as the specification of each its field (unit, label, short textual description, reference to documentation, and so on).
This auxiliary information is used for proper visualisation of the data.

The end user receives the information from the distance catalogue tables gathered together on-the-fly in a single view.
This approach allows us, on the one hand, to avoid redundancy of the stored data and, therefore, to avoid the risk of inconsistency, 
and, on the other hand, to improve the readability of information by minimizing references to a different tables.

\subsubsection{Measurement table}

The measurement table contains the following fields:

\begin{description}
\item[\textsc{pgc}] 
  is the Principal Galaxies Catalogue number, which was invented by \citet{PGC}.
  We use the standard identification schema of the HyperLEDA database.
  Each object has a unique number, used to link the data from the different catalogues, and, in particular, to its various designations.
  When a `search by name' is performed, the system first resolves the provided name into a \textsc{pgc} number, 
  which in turn is used to access the data.

\item[\textsc{modulus}, \textsc{e\_modulus}] 
  contains a published distance modulus and its one-sigma measurement error in mag.
  It is a so-called true distance modulus, $(m-M)_0$, 
  which is corrected for both the extinction in the our Galaxy and the absorption in the host galaxy.
  The web interface also shows the distance in the linear scale in units of Mpc.

\item[\textsc{quality}]
  describes the quality of the data. 
  It is divided in the two parts. 
  The first one is a data-related set of flags: `uncertain' (:), `preliminary' (p), `low-limit' ($>$), or `compilation' (c).
  They are based on information provided by authors. 
  The `uncertain' flag reduces the weight of the measurement by a factor 2 during the homogenization procedure.
  Other flags lead to elimination of the measurement from averaging if other data are available.
  The `reject' (!) code describes our HyperLEDA knowledge about the reliability of the data and 
  discards erroneous measurements from the homogenization procedure.

\item[\textsc{iref}] 
  is an internal code for the publication.
  It points toward the general bibliographic table of HyperLEDA, 
  which associates it with the ADS/CDS \textsc{bibcode}, when available, 
  as well as with a short description of the article, including the first author and the year of the publication. 
  This allows us to search the whole database by author or by a given paper.

\item[\textsc{method}] 
  codes the distance determination method.
  The list of methods is given in Sect.~\ref{sect:method}

\item[\textsc{calib}]
  points out a detailed information about the calibration relation used for a measurement.
  This field allows us to group the data obtained with the same calibration.

\end{description}

In addition to these fields, which are exposed to the users of the database, some additional fields are kept for internal use.

\begin{description}

\item[\textsc{origname}]
  is the name of the object as cited in the original reference. 
  This allows us to trace a given measurement back to the source table.

\item[\textsc{note}] 
  contains important remarks on specific measurement. 

\end{description}

Finally, some fields exposed to the user are derived from an on-the-fly calculation or are computed during the homogenization.

\begin{description}

\item[\textsc{distance}]
is the linear distance, in Mpc, computed from \textsc{modulus}. 
It is given for convenience.

\item[\textsc{modc}]
is the homogenized distance modulus
after the calibration correction is applied (see Sect.~\ref{sect:homogenization}). 
The best distance determination for a galaxy can be derived from a weighted average of the individual \textsc{modc}.

\end{description}

\subsubsection{Calibration table}

The field \textsc{calib} links the measurement to the calibration table with the following fields:

\begin{description}

\item[\textsc{calib}] 
  is a unique code used to identify a specific distance calibration.

\item[\textsc{method}] 
  is a distance determination method as in the measurement table.

\item[\textsc{iref}] 
  is an internal bibliographical code of the article where the calibration is published. 

\item[\textsc{note}] 
  contains miscellaneous remarks useful for understanding a given calibration.

\end{description}

\subsection{Distance determination methods}
\label{sect:method}

To date, a number of distance determination methods have been invented.
They vary in the class of objects used (e.g.\ Cepheids) and in the physical background (e.g.\ period-luminosity relation).
The corresponding information is coded in the \textsc{method} field of the measurement table.
Below we briefly describe of these methods.
We divide the list into three parts: 
(i) direct distance determination methods,
(ii) use of specific stellar objects as a standard rule or a standard candle,
and (iii) techniques based on scaling laws for galaxies.

\textit{Direct methods} determine distances straight from the measurement data and do not depend on external calibrations.
They are the basis for constructing the cosmic distance ladder.
The most important distance estimates use the trigonometric parallax of individual stars.
The methods of statistical parallax and moving cluster parallax allow us to derive distances for groups of stars.
It is very useful for calibrating methods based on the luminosity of the Cepheids and RR Lyrae.
Unfortunately, these methods are usually restricted to our Galaxy or to its nearby satellites.
A notable exception is NGC\,4258, whose precise `maser' distance of 7.6\,Mpc (see hereafter) is precious for calibrating the other methods.

\begin{description}

\item[DEB:]
  The detached eclipsing binaries (DEB) provide an accurate geometric method for distance determination.
  The fundamental parameters of the stars (the radii, effective temperatures, masses, and luminosities)
  can be determined from the light and radial velocity curves of an eclipsing binary.
  This method is independent of any intermediate calibration steps.

\item[EPM, ESM:]
  The expanding photosphere method (EPM) and the expanding shock-front method (ESM)
  are a geometric distance determination technique based on comparing radial velocities with proper motion 
  of an expanding shell after a supernova explosion.

\item[Maser:] 
  The method is based on studying kinematics of an accretion disk around a supermassive black hole by a radio maser emission.
  It gives a direct geometric estimate of an absolute distance.
  \citet{HRM2013} measured the distance of 7.6\,Mpc with 3\,\% uncertainty to the Seyfert II galaxy NGC\,4258 
  using ten years observations of the H$_2$O maser.

\end{description}

\noindent
A wide number of methods uses \textit{individual objects or stellar populations} in the galaxies for distance determination.
This class contains some of the most precise and important distance indicators for extragalactic astronomy: 
the Cepheids and RR Lyrae variable stars, 
the tip of the red giant branch (TRGB) and the horizontal branch (HB) stars.
These distance indicators can be calibrated using the direct methods described above\footnote{
The references cited in the description of each method are examples of recent studies. 
A wider overview can be obtain in the body of the catalogue.
}.
Except for SN\,Ia, all these methods are only effective for the nearby Universe on a scale from several to a few dozen Mpc.

\begin{description}
\item[BBSLF, BRSLF, BS, BS3B, BS3R:] 
  All these methods use the luminosity of the brightest stars in galaxies as standard candles. 
  BBSLF and BRSLF consider the luminosity function of the brightest blue and red stars. 
  BS3B and BS3R take the mean absolute magnitude of the three brightest blue or red stars.
  The luminosity of the brightest blue and red supergiants depends on the magnitude of the parent galaxy \citep{RRR1994}.  


\item[Cepheids:] 
  This is one of the most important standard candles.
  The method is based on the period-luminosity (PL) relation for Cepheid variable stars.
  There are many calibrations of the relation in different pass-bands using
  the Galactic or Large Magellanic Cloud (LMC) PL relation, for example,
  the Hubble Space Telescope Key Project On the Extragalactic Distance Scale \citep{HSTKP},
  the HIPPARCOS trigonometric parallaxes \citep{FC1997},
  or the Baade-Wesselink methods \citep{S+2011}.
  The calibration will be dramatically improved in the coming years thanks to the GAIA astrometric satellite, which starts to operate now.

\item[CMD:]
  This uses various features of the composite colour-magnitude diagram (CMD) of a galaxy resolved into individual stars 
  to estimate the distance by comparison with template CMD or with theoretical isochrones. 
  For instance, \citet{D2000} developed the software that fits the observed CMD with synthetic data
  to simultaneously estimate the distance and the star formation history of a galaxy.

\item[CS:]
  The carbon-rich stars (CS) in the TP-AGB phase form the horizontal red tail on the CMD, and are about 0.5 mag brighter than the TRGB.
  \citet{BD2005} found the absolute $I$-band magnitude of CS as a function of the metallicity of the parent galaxy:
  $\langle M_I\rangle=-4.33+0.28\FeH$.

\item[FGLR:] 
  The flux-weighted gravity-luminosity relationship (FGLR)
  is a technique to derive the distance from a spectral analysis of the B and A supergiant stars \citep{FGLR}.
  It is based on a tight correlation between the absolute bolometric magnitude and the flux-weighted gravity, $g/T^4_\mathrm{eff}$.

\item[GCLF:] 
  The old globular cluster luminosity function (GCLF) method uses 
  the peak (or turnover, TO) of the GCLF as a standard candle.
  For instance, \citet{D+2006} derived $M_{V,\mathrm{TO}}=-7.66\pm0.09$ with an adopted LMC distance modulus of 18.50.

\item[GCR:]
  The median of the globular cluster half-light radii (GCR) of $2.7\pm0.3$ pc \citep[for example]{J+2005} 
  can be used as a standard ruler for the distance estimate.
  The half-light radius of individual GC needs to be corrected for colour, surface brightness, and host galaxy colour.

\item[HB, BHB:]
  These methods use the horizontal branch (HB) or the blue horizontal branch (BHB) stars as standard candles.
  \citet{Carretta+2000} reported the relation between absolute magnitude and metallicity of HB:
  $M_V(\mathrm{HB}) = (0.13\pm0.09)(\FeH+1.5) + (0.54\pm0.07)$.

\item[MS:]
  This method fits the position of the main sequence below the turn-off with theoretical isochrones or with template CMD.
  It is related to the CMD distance determination method.

\item[Miras:] 
  Mira Ceti stars are long-period variable stars in the asymptotic giant branch phase.
  \citet{IM2011}, among others, derived the period-magnitude relations for Mira-like variables in the LMC
  using bolometric, near- and mid-infrared magnitudes.

\item[PNLF:]
  This method uses the sharp exponential truncation of the planetary nebulae luminosity function (PNLF) as a standard candle.
  The zero-point, $M^*=-4.48$, is based on the M\,31 distance of 710 kpc \citep{C+2002}.

\item[RC:] 
  The red clump (RC) is populated by core helium-burning stars of intermediate age.
  Their mean absolute magnitude provides a standard candle for distance determination. 
  \citet{GS2001} found important non-linear dependences on both the age and the metallicity of the stellar population.

\item[RRLyrae:]
  The method is based on the mean absolute magnitude for RR Lyrae variable stars, which depends on metallicity:
  $M_V(\mathrm{HB}) = (0.18\pm0.09)(\FeH+1.5) + (0.57\pm0.07)$ \citep{Carretta+2000}.

\item[RSV:]
  This method uses the period-luminosity relation for the red supergiant variable (RSV) stars.
  The calibration of the PL relation by \citet{PJC2000} adopts the distance modulus of 18.50 mag for LMC.
  The RSVs as well as the Miras are long-period variable stars.

\item[SBF:] 
  The surface brightness fluctuations (SBF) method relies on the luminosity fluctuations 
  that arise from the counting statistics of stars that contribute to the flux in each pixel of an image \citep{SBF}.
  The absolute fluctuation magnitude depends on the stellar populations and, consequently, on the colour of the galaxy.
  It can only be applied to old stellar populations.

\item[SNIa:] 
  Because of their extremely high luminosity and regular behaviour, the type Ia supernovae (SN\,Ia) provide 
  a powerful tool for measuring cosmological distances.
  The method uses the relationship between the light-curve shape and the maximum luminosity of a SN\,Ia.

\item[TRGB:] 
  The tip of the red giant branch (TRGB) is an excellent distance indicator for nearby galaxies that are resolved into individual stars. 
  The method, relying on the old stellar population, can be used for galaxies of any morphological types. 
  Thanks to the shallow colour-dependence of the magnitude of the TRGB in the $I$ band, 
  the method is one of the most precise distance indicators.
  For example, \citet{TRGB2} calibrated the zero-point of the TRGB method using HB stars:
  $M_{I}^\mathrm{JC} = -4.05(\pm0.02) + 0.22(\pm0.01)[(V-I)-1.6]$.

\end{description}

\noindent
Methods that are based on \textit{scaling relations} are empirical relationships between 
the intrinsic luminosity of a galaxy and its properties such as kinematics, and surface brightness.
The most important ones are the Tully-Fisher (TF) relation for spirals and the fundamental plane (FP) for early-type galaxies.
Because the methods use the total luminosity of a galaxy as a standard candle, they can be applied on scales of up to several hundred Mpc.
These methods provide only a low precision for individual measurements, but they give good results in a statistical sense with huge sets of data.
This is especially true for the Tully-Fisher relation, where obtaining observational data is relatively inexpensive.
The TF and FP methods allow us to investigate the cosmic flows in the Universe on scales of several hundred Mpc \citep{CosmicFlows2}.

\begin{description}
\item[FJ:]
  The Faber-Jackson (FJ) relation provides a standard candle for elliptical and early-type galaxies 
  based on the relationship between absolute magnitude and central velocity dispersion.

\item[FP:]
  The fundamental plane (FP) is a distance determination method for early-type galaxies 
  based on relation between the absolute magnitude, effective radius, velocity dispersion, and mean surface brightness.
  $\log D = \log r_\mathrm{e} -1.24\log\sigma +0.82\log\langle I\rangle_\mathrm{e} +0.173$ \citep{K+2000}.

\item[SB-M, Sersic-M:]
  The methods using the surface brightness-total magnitude relation (SB-M) or the Sersic index-total magnitude relation (Sersic-M)
  can be considered as a rough distance estimate for small-mass elliptical galaxies.

\item[Sosie:]
  The method of `look-alike' (sosie in French) was proposed by \citet{P1984}.
  It is based on the idea that galaxies with the same morphological type, the same inclination, 
  and the same \HI{} line width must have the same absolute luminosity according to the TF relation.

\item[TF, BTF:]
  The Tully-Fisher (TF) method is a standard candle based on empirical relationship between the absolute magnitude of a spiral galaxy 
  and its maximum rotational velocity, estimated by a \HI{} line width.
  The recent calibration of $I$-band TF relation gives
  $M_I^{b,i,k} = -21.39-8.81(\log W^i_\mathrm{mx}-2.5)$ \citep{TC2012}.

  The baryonic Tully-Fisher (BTF) relation uses the relationship between the amplitude of rotation and the baryonic mass of the galaxy.
  This relation takes into account not only the stellar light from optical data as in the original TF relation, 
  but also the mass of gas in neutral and molecular forms. 
  The BTF relation is similar to the TF for giant spiral galaxies, 
  and it represents an improvement for dwarf galaxies with circular velocities below 90 \kms{} \citep{M+2000},
  where cold gas represents an important and variable dynamical component. 
  The BTF can also be applied to gas-rich dwarf elliptical galaxies \citep{dR+2007}.

\end{description}

\subsection{Content of the catalogue}

\begin{figure}
\centerline{
\includegraphics[width=0.45\textwidth,clip]{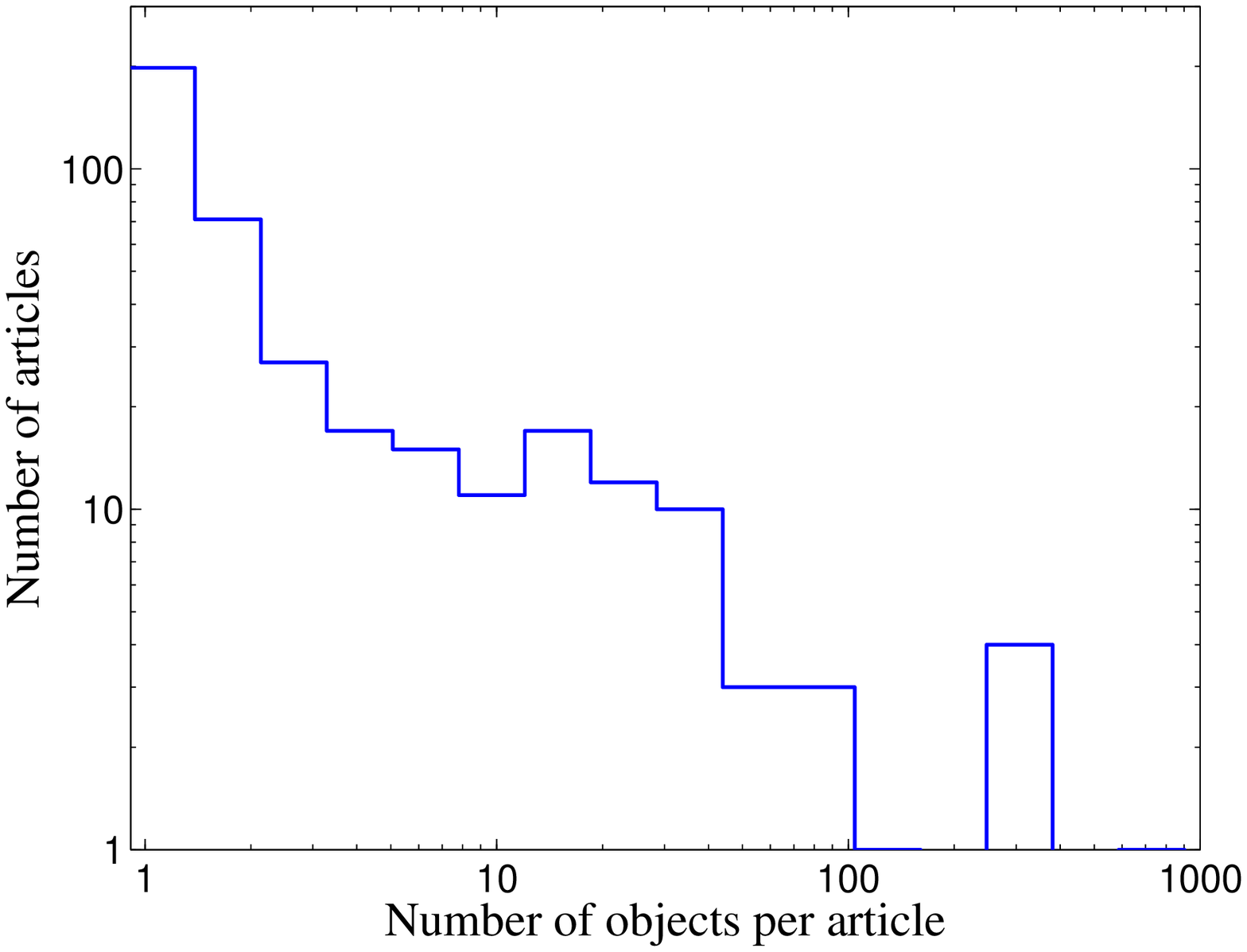}
}
\caption{
Distribution of articles by the number of published galaxies for the distance catalogue.
}
\label{f:obj-art}
\end{figure}

\begin{figure*}
\centerline{
\includegraphics[width=0.95\textwidth,clip]{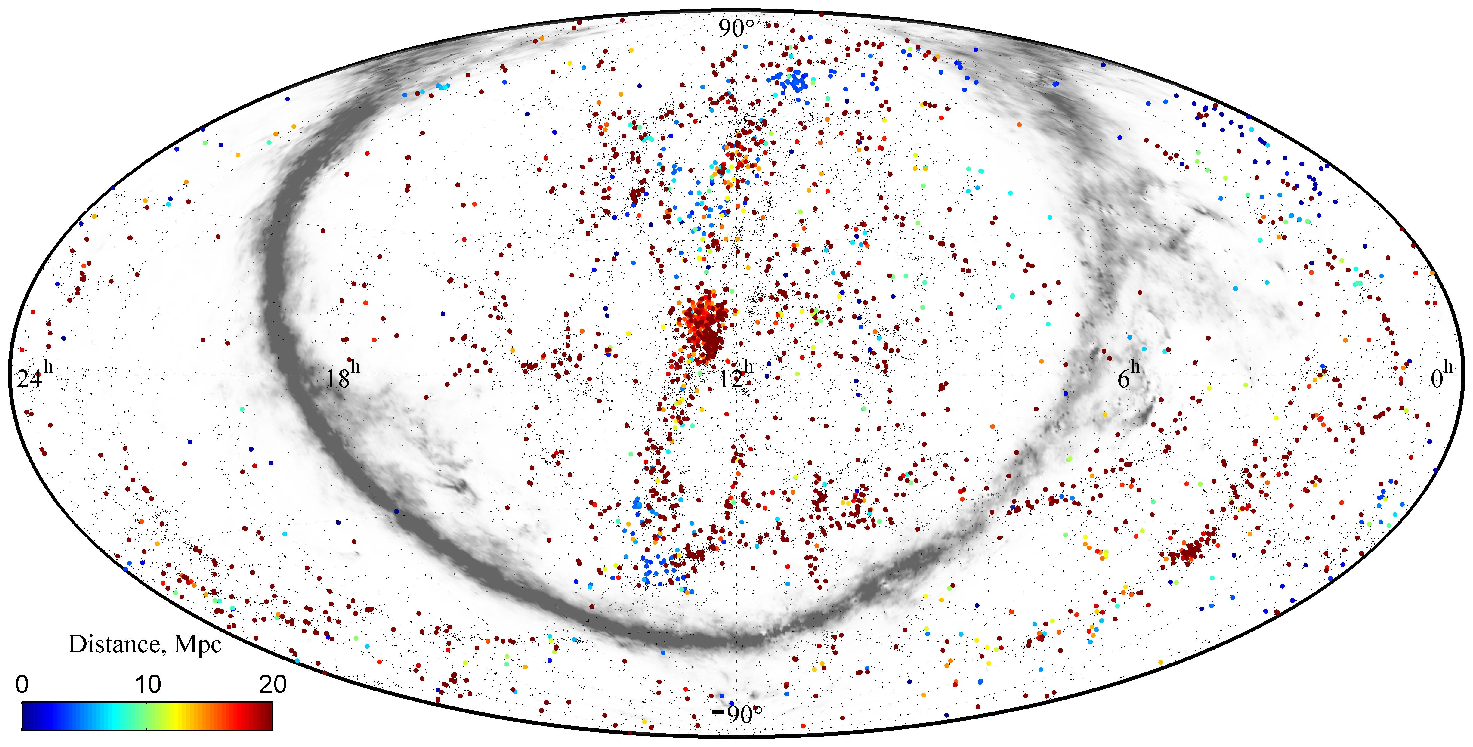}
}
\centerline{
\includegraphics[width=0.95\textwidth,clip]{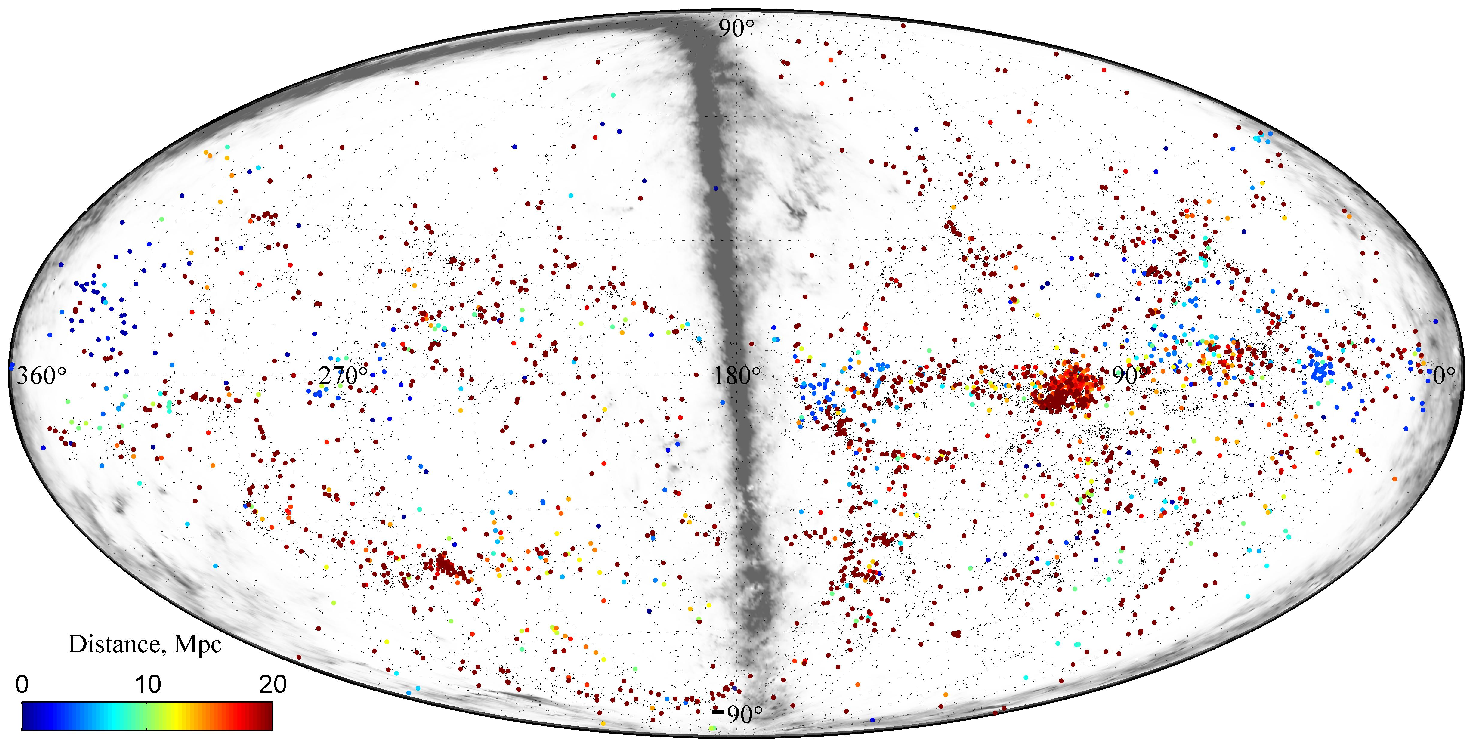}
}
\caption{
All-sky distribution of galaxies with known redshift-independent distance estimate in equatorial (top panel)
and in supergalactic coordinates (bottom panel).
The distance is shown by colour from nearby (blue) to distant (brown).
The small black dots represent the distribution of galaxies on a scale of 40\,Mpc.
The zone of high extinction in the Milky Way is indicated in grey.
}
\label{f:sky20}
\end{figure*}

The distance catalogue currently contains \catRec{} distance measurements for \catObj{} galaxies compiled from \catArt{} articles. 
We pay special attention to the Local Volume, $D\lesssim10$\,Mpc, 
because the vast majority of high-precision distances are measured for nearby galaxies.
For instance, at the moment, we gather 2\,594 distance moduli
for 492 objects in this local part of the Universe,
in particular, 1\,338 estimates
for 347 galaxies based on the TRGB method.
As it is illustrated in Fig.~\ref{f:obj-art}, most of the articles list measurements for a single or a few galaxies. 
The distance catalogue describes the specifics of the work on determining the distances to galaxies.
Single galaxies are presented in 198 articles (50\,\%) and only 13 articles (3\,\%) give measurements for more than 50 galaxies. 

The distribution on the sky of the galaxies with known redshift-independent distances is shown in Fig.~\ref{f:sky20}.
The clumpy grey belt represents the regions of strong extinction in our Galaxy (the Zone of Avoidance). 
The colour of the dots codes the distance to a galaxy, from blue for nearby objects to brown for the distant ones.
Galaxies farther away than 20\,Mpc are shown as brown filled circles.
The small black dots represent all galaxies with $V_\mathrm{LG} < 3\,500$ \kms{} (i.e.\ within 50\,Mpc),
roughly encompassing the Local Supercluster, 
whose core, the Virgo cluster, lies near the centre of the top-panel map 
(about $\mathrm{RA}=12\fh5$, $\mathrm{Dec}=12\degr$).
Most of the galaxies are concentrated near the supergalactic plane, 
revealing the flat shape of the Local Supercluster,
which is clearly seen in the supergalactic coordinates in the bottom panel of Fig.~\ref{f:sky20}.
Fortunately, the Virgo cluster is located near the Galactic north pole,
and the sheet of the Local Supercluster galaxies is perpendicular to the plane of the Milky Way.
Because the Local Group stands on the edge of the Local Supercluster,
we do not see its extension in the direction opposite to the Virgo cluster.

\begin{figure}
\centerline{
\includegraphics[width=0.45\textwidth,clip]{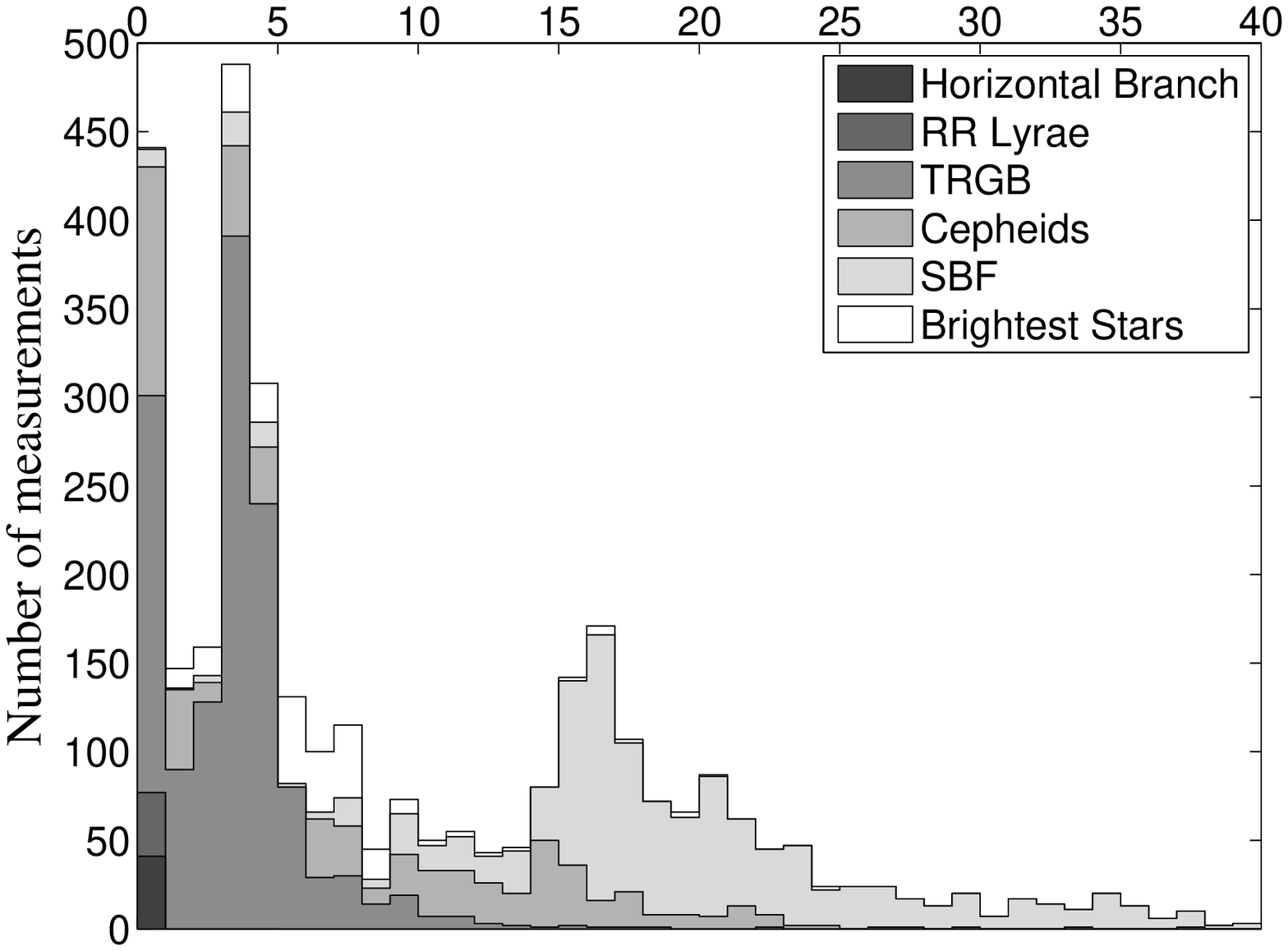}
}
\centerline{
\includegraphics[width=0.45\textwidth,clip]{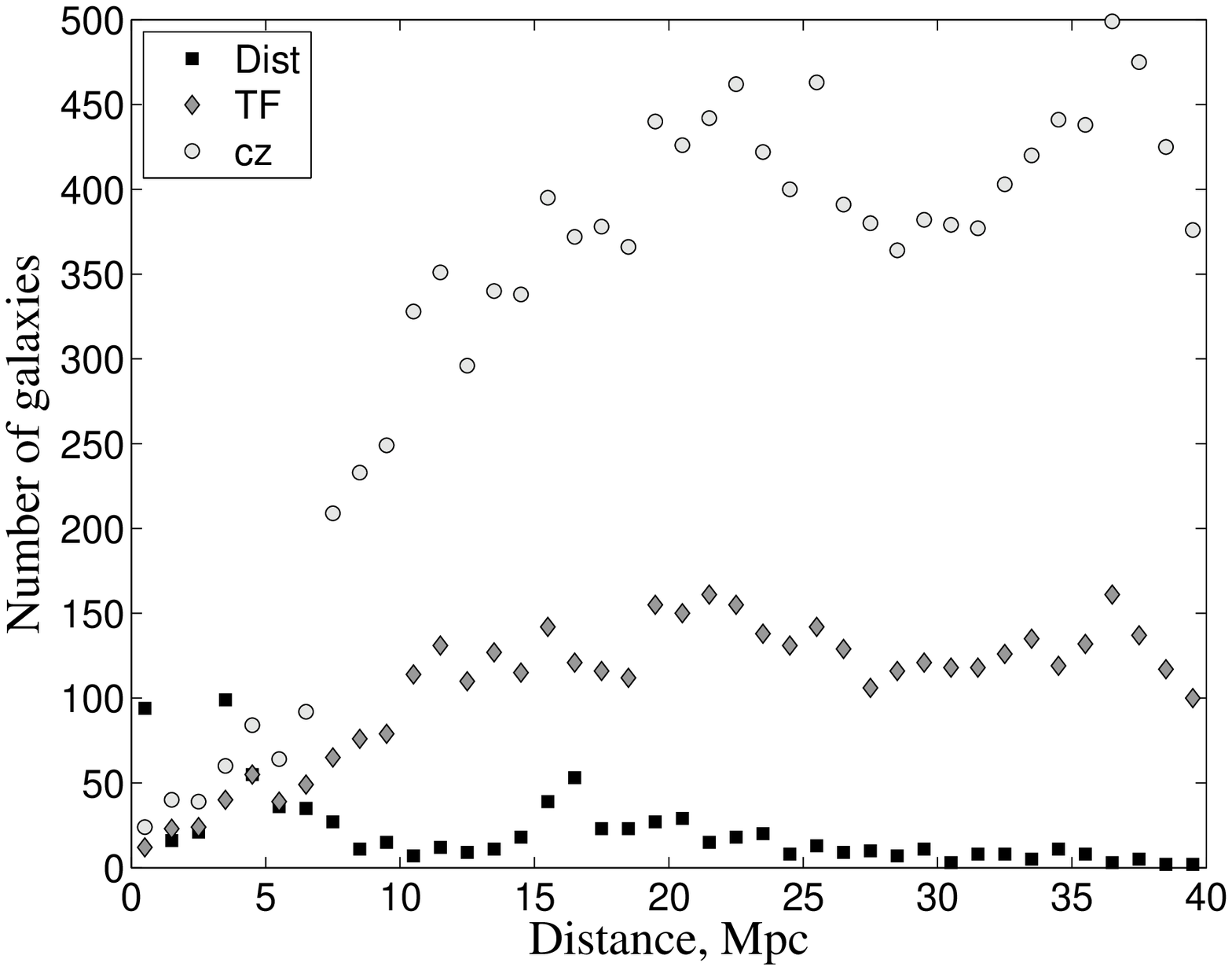}
}
\caption{
Top panel: distribution of the individual distance estimates versus distance.
Only the five most significant methods, 
namely
the horizontal branch, 
the RR Lyrae, 
the tip of the red giant branch (TRGB), 
the Cepheids, 
the surface brightness fluctuations (SBF), 
and the brightest stars,
were considered to build the histogram.
While the outer envelope shows the total number of measurements in each 1\,Mpc bin, the different grey tones distinguish the contribution of each method.
The bottom panel displays the distribution of the individual galaxies.
The black squares correspond to objects with known redshift-independent distance estimate (Dist), excluding the scaling relations.
The distributions of galaxies with Tully-Fisher data (TF) is shown by grey diamonds,
and galaxies with known redshift, $cz$, are illustrated by open circles.
TF is computed using the \HI{} and photometric parameters of HyperLEDA,
while the redshift is transformed into distance using the Hubble law.
}
\label{f:dist}
\end{figure}

Figure~\ref{f:dist} illustrates the distribution of individual measurements by different methods as a function of the distance.
The various shades of grey in the top panel distinguish the contribution of the most significant methods, 
namely the horizontal branch, the RR Lyrae, the tip of the red giant branch, the Cepheids, 
the surface brightness fluctuations, and the brightest stars. 
The first peak below 1\,Mpc corresponds to the Local Group galaxies.
The second spike around 4--5\,Mpc marks the systematic study of the Local Volume \citep{UNGC},
where the most prominent groups of galaxies lie around Cen\,A, M\,81, IC\,342 and 
several other giant galaxies are situated about 3--4\,Mpc away from the Milky Way. 
\citet{McCall2014} named this remarkable distribution `the Council of Giants'.
Beyond 5\,Mpc, the sharp cut-off in the number of measurements reflects 
the increasing observational difficulty and cost of TRGB determinations. 
To reach the necessary limiting magnitude, several HST orbits are required for a single field.
The third maximum, around 16--17\,Mpc, has a physical origin.
It corresponds to the centre of the Virgo cluster.
The figure shows that the prevalence of the different indicators changes with the distance. 
HB and RR Lyrae stars can be used only inside the Local Group, below 1\,Mpc.
The tip of RGB is much brighter, $M_I=-4$, and the method is commonly used up to 10\,Mpc, 
with extension up to about 20\,Mpc in particular cases.
The use of Cepheids is much more expensive in terms of observations, because it requires time series.
This explains the relatively small number of measurements, although it is one of the most precise and refined methods.
Currently, the Cepheids allow us to confidently derive the distances up to 20\,Mpc.
The SBF method is not as precise as the previous ones, 
but an accuracy about 0.2 mag in the distance modulus can be achieved for a wide range of distances up to 40--50\,Mpc.
The brightest-stars method cannot be considered as precise, 
but in many cases, it is the only possibility to estimate the distance between 5 to 10\,Mpc.
Farther out, the method based on the TF relation is the most effective.
Recent calibrations give a scatter of about 0.34--0.38 in the different band-passes \citep{TP2000}.
HyperLEDA contains \HI{} line widths for about 37\,000 galaxies.
Taking into account the existence of photometry and limiting the inclination angle more than $45\degr$, 
this allows us to estimate a TF distance for about 19\,000 galaxies on a scale of several hundred Mpc.
The distribution of the galaxies with available data for TF distance estimate is shown 
in the bottom panel of Fig.~\ref{f:dist} by grey diamonds for comparison.
A similar distribution of galaxies with known redshift is displayed by open circles.
This means that for large scales the TF distance estimate is possible only for a third of the known galaxies.
In addition, black squares show the objects with precise distance measurements.
It is clear that our knowledge on accurate distances is more or less complete only in the Local Volume on a scale of up to 5--7\,Mpc.

\section{Distance homogenization}
\label{sect:homogenization}

\begin{table*}
\caption{Comparison of different sets of distance determinations with the TRGB set from EDD.}
\label{table:calibset}
\begin{tabular}{clrl@{$\;\;\pm$}lrl}
\hline\hline
& sample   & N  & \multicolumn{2}{c}{$\langle\mu-\mu_\mathrm{EDD}\rangle$} & $\sigma$ & notes \\
\hline
\multicolumn{6}{l}{TRGB} \\
& \citet{TRGB2} excluding EDD   &  83 & $-0.016$ & 0.008 & 0.060 & \citet{TRGB2} \\
& \citet{Makarov+2013}          &  26 & $-0.022$ & 0.005 & 0.025 & \citet{TRGB2} \\
& \citet{Radburn-Smith+2011}    &  13 & $-0.021$ & 0.024 & 0.086 & \citet{TRGB2} \\
& \citet{Dalcanton+2009}        &  87 & $-0.064$ & 0.008 & 0.062 & Padova isochrones \\
& $M_I(\mathrm{TRGB})=-4.05$    & 181 & $-0.027$ & 0.006 & 0.062 & \\
& \citet{Lee+1993}              &  30 & $-0.073$ & 0.021 & 0.114 & \\
\multicolumn{6}{l}{Cepheids} \\
& \citet{Saha+2006}             &  15 & $-0.013$ & 0.048 & 0.157 &  \\
& \citet{Fouque+2003}           &   7 & $+0.002$ & 0.041 & 0.103 &  \\
& \citet{Kanbur+2003}           &   8 & $+0.003$ & 0.062 & 0.197 &  \\
& \citet{Tammann+2003}          &   7 & $+0.063$ & 0.047 & 0.110 &  \\
& \citet{HSTKP}                 &  13 & $-0.065$ & 0.030 & 0.100 &  \\
& \citet{HSTKP}                 &  14 & $-0.016$ & 0.034 & 0.143 & metallicity corrected \\
& \citet{Lanoix+1999}           &  16 & $-0.029$ & 0.044 & 0.147 &  \\
& \citet{Udalski+1999}          &   8 & $+0.015$ & 0.061 & 0.196 & corrected to $\mu_\mathrm{LMC}=18.50$ \\
& \citet{Gieren+1998}           &   7 & $-0.005$ & 0.047 & 0.109 &  \\
& \citet{Madore+1991}           &  13 & $+0.007$ & 0.035 & 0.117 &  \\
\multicolumn{6}{l}{SBF} \\
& \citet{Jensen+2003}           &   8 & $-0.223$ & 0.076 & 0.223 &  \\
& \citet{Tonry+2000}            &  16 & $+0.092$ & 0.057 & 0.250 &  \\
\multicolumn{6}{l}{PNLF} \\
& \citet{Ciardullo+1989}        &  10 & $+0.043$ & 0.089 & 0.225 &  \\
\multicolumn{6}{l}{Brightest stars} \\
& Brightest blue stars          &  61 & $-0.039$ & 0.076 & 0.586 &  \\
& Brightest red stars           &  21 & $-0.080$ & 0.140 & 0.632 &  \\
\hline\hline
\end{tabular}
\end{table*}

The measurements collected in this catalogue form an inhomogeneous data set
where each individual publication uses a specific distance scale and is affected by its own systematics. 
As for the other characteristics of galaxies, HyperLEDA aims to compensate these discrepancies.
The homogenization in HyperLEDA attempts to reduce these effects as much as possible. 

The strategy adopted in the present version of the catalogue is the following:
First, we define a set of calibrators that determines our distance scale. 
Then, we apply zero-point correction accounts for the shift between a given calibration system and our adopted distance scale.

These different steps are described below.

\subsection{Distance standards}

The goal is to define a set of galaxies with precise distance measurements that can be used to calibrate other systems. 
This will define our own distance scale.

The most recent and widely adopted distance scale is based on Cepheid measurements in 31 galaxies reported in the HST key project \citep{HSTKP}.
The quality of the data are still not surpassed, but the small size of the sample limits its usage for calibrating other methods and datasets.
To circumvent this limitation, we can consider building our calibration set around the `CMDs/TRGB' catalogue \citep{EDD:CMD} of EDD, 
encompassing highly homogenized measurements for 305 galaxies
using the maximum-likelihood TRGB fit by \citet{TRGB1} and the TRGB zero-point by \citet{TRGB2}.

Table~\ref{table:calibset} presents a comparison of different data samples with the EDD reference sample.
Column 1 gives the reference of the datasets.
Column 2 is the number of common objects compared with our EDD sample.
Columns 3 and 4 present the mean bias and its one-sigma uncertainty.
Column 5 shows the dispersion around the bias.
Column 6 gives an additional description of the compared measurements.
The datasets are grouped by different distance determination methods.

The table gives a first taste of the intrinsic precision of individual methods.
The TRGB and Cepheids distances have a precision of about 0.1 mag.
The SBF and PNLF methods are precise to 0.2 mag.
Finally, brightest stars are only precise to 0.6 mag.
Therefore, this method is useful only when no other measurement available.

While the TRGB method displays a very low internal dispersion (0.06 mag), the various measurement sets are not fully independent 
because they are often based on the same original images and differ only by some aspects of the analysis and calibration.
Low dispersion indicates that the photometric errors probably outweigh the systematics associated with the different techniques of tip detection.

The comparisons gathered in Table~\ref{table:calibset} reveal a fair consistency, generally within two sigma.
Nevertheless, the small $-0.02$ mag bias between any of the considered TRGB sets and the `CMDs/TRGB' sample appears to be statistically significant. 
Its origin is not clear.
A comparison of the TRGB method with various period-luminosity relations for Cepheids shows that they agree excellently.
In most cases, the average difference does not exceed 0.02 mag. 
In other words, the TRGB and Cepheids distance scales are consistent on a level better than 1\,\%.
Note that the TRGB calibration by \citet{TRGB2} is based on the luminosity of the horizontal branch reported by \citet{Carretta+2000} and 
is independent of Cepheids distances.
Although the SBF method is calibrated on the Cepheid variables it requires a large correction to be consistent with other methods.

In summary, the CMDs/TRGB catalogue \citep{EDD:CMD} can be considered as the basis for the standard sample of galaxies for the distance homogenization.
Many authors point out that the TRGB method has a precision similar to that of the Cepheids \citep{Sakai+1996,Bellazzini+2001}.
\citet{TRGB2} found an excellent agreement between TRGB and Cepheid scales ($\mu_\mathrm{Ceph}-\mu_\mathrm{TRGB} = -0.01\pm0.03$) using 15 nearby galaxies.
Our analysis confirms this conclusion.
We selected 194 galaxies with an uncertainty better than $0.1$ mag from the CMDs/TRGB catalogue to be the core of our calibration set.
This sample covers nearby galaxies well, but is still limited for calibrations of long-distance methods.
As can be seen from Fig.\,\ref{f:dist}, current TRGB measurements are mostly bounded by a distance of 5--6\,Mpc.
This restriction is especially strong for early-type galaxies because the Local Volume is almost free of giant ellipticals.
To partially palliate this problem we adopted 29 measurements with an uncertainty better than $0.12$ mag 
reported \citet[corrected for metallicity effect]{HSTKP} from the final results of the Hubble Space Telescope key project.
It is interesting to note that excluding two galaxies, NGC\,5253 and IC\,1613 with measurement errors of 0.14 and 0.15 mag,
makes the accordance between TRGB and Cepheids scales ideal, $\langle\mu_\mathrm{KP}-\mu_\mathrm{EDD}\rangle_w = -0.001 \pm 0.034$.
This dataset effectively extends to 15--20\,Mpc, favouring spiral galaxies.
We include a new geometric maser distance, $D=7.6$\,Mpc or $\mu=29.40$ mag, to NGC\,4258 \citep{HRM2013}, 
which agree excellently with the TRGB estimate $\mu_\mathrm{TRGB}=29.42\pm0.04$ \citep{EDD:CMD} and
with the \citet{HSTKP} value of $\mu_\mathrm{Ceph}=29.44\pm0.07$ mag.
Our sample of the standard distance measurements contains 211 galaxies.

\subsection{Calibration correction}

Most of the data in the catalogue refer to the calibration used to determine the distance.
It allows us to group measurements made with the same method and the same calibration. 
For each calibration, we determine the mean bias as the weighted average of the difference between individual measurements and our standard distance set. 
We apply this correction to data if an intersection with the standard distance set has at least five objects. 
Now, the bias is estimated for 58 calibrations of 13 distance determination methods.
Field \textsc{modc} shows the homogenized distance modulus for individual measurements after the calibration correction is applied.

\section{Conclusion}
\label{sect:conclusion}

Over the past decade, the number of available data in extragalactic astronomy has increased enormously, thanks to modern surveys. 
However, despite the increase in quality of the measurement, when a sample for calibrating of distance indicators needs to be established, 
it remains necessary to critically review a large number of various sources of measurements. 
The systematics that affect specific methods or specific series of measurements have to be clearly understood, 
and the need for well-documented compilations remains as acute as ever.
Fortunately, the access and the tools to handle the published data are improving, which keeps the effort at a sustainable level. 

We here presented the compilation catalogue of redshift-independent distances in the HyperLEDA database. 
Currently, we have surveyed \catArt{} publications that yield \catRec{} distance estimates for \catObj{} galaxies.
Most of them are concentrated in the Local Volume.
Each individual data series is recalibrated to a common distance scale based on a carefully selected set of high-quality measurements.
In addition, the large collections of \HI{} and photometric data in HyperLEDA enable distance estimates for 19\,000 spirals using the TF relation.

The distance catalogue is available through the web-interface of the HyperLEDA database (\url{http://leda.univ-lyon1.fr}), 
which is currently maintained by the Observatoire de Lyon (France) and 
the Special Astrophysical Observatory of the Russian Academy of Sciences (Russia).

\begin{acknowledgements}

We are thankful to Georges Paturel for a long-standing support and interest in the project,
and to Chantal Petit for years of work on HyperLEDA.
We appreciate the help and contribution in software development of 
Rumen Bogdanovski, Sergey Koposov, Vladimir Georgiev, Igor Chilingarian, and Ivan Zolotukhin.
We are grateful for catalogue support to Anatoly Zasov, Valeri Golev, and Natasa Gavrilovic.
We thank Mina Koleva and Galina Korotkova for data mining, analysis, and processing.
We thank the numerous users who report missing or erroneous data, make suggestions to improve the service, and provide encouragement.
DM acknowledges the support of the Russian Science Foundation grant 14--12--00965.
DM thanks Observatoire de Lyon for hospitality during the work.

\end{acknowledgements}

\bibliographystyle{aa}
\bibliography{leda3}   

\begin{thebibliography}{80}
\expandafter\ifx\csname natexlab\endcsname\relax\def\natexlab#1{#1}\fi

\bibitem[{{Ahn} {et~al.}(2013){Ahn}, {Alexandroff}, {Allende Prieto}, {Anders},
  {Anderson}, {Anderton}, {Andrews}, {Aubourg}, {Bailey}, {Bastien}, \&
  et~al.}]{SDSSdr10}
{Ahn}, C.~P., {Alexandroff}, R., {Allende Prieto}, C., {et~al.} 2013, ArXiv
  e-prints

\bibitem[{{Baillard} {et~al.}(2011){Baillard}, {Bertin}, {de Lapparent},
  {Fouqu{\'e}}, {Arnouts}, {Mellier}, {Pell{\'o}}, {Leborgne}, {Prugniel},
  {Makarov}, {Makarova}, {McCracken}, {Bijaoui}, \& {Tasca}}]{EFIGI}
{Baillard}, A., {Bertin}, E., {de Lapparent}, V., {et~al.} 2011, \aap, 532, A74

\bibitem[{{Battinelli} \& {Demers}(2005)}]{BD2005}
{Battinelli}, P. \& {Demers}, S. 2005, \aap, 442, 159

\bibitem[{{Bellazzini} {et~al.}(2001){Bellazzini}, {Ferraro}, \&
  {Pancino}}]{Bellazzini+2001}
{Bellazzini}, M., {Ferraro}, F.~R., \& {Pancino}, E. 2001, \apj, 556, 635

\bibitem[{{Bonnarel} {et~al.}(2000){Bonnarel}, {Fernique}, {Bienaym{\'e}},
  {Egret}, {Genova}, {Louys}, {Ochsenbein}, {Wenger}, \& {Bartlett}}]{Aladin}
{Bonnarel}, F., {Fernique}, P., {Bienaym{\'e}}, O., {et~al.} 2000, \aaps, 143,
  33

\bibitem[{{Carretta} {et~al.}(2000){Carretta}, {Gratton}, {Clementini}, \&
  {Fusi Pecci}}]{Carretta+2000}
{Carretta}, E., {Gratton}, R.~G., {Clementini}, G., \& {Fusi Pecci}, F. 2000,
  \apj, 533, 215

\bibitem[{{Ciardullo} {et~al.}(2002){Ciardullo}, {Feldmeier}, {Jacoby}, {Kuzio
  de Naray}, {Laychak}, \& {Durrell}}]{C+2002}
{Ciardullo}, R., {Feldmeier}, J.~J., {Jacoby}, G.~H., {et~al.} 2002, \apj, 577,
  31

\bibitem[{{Ciardullo} {et~al.}(1989){Ciardullo}, {Jacoby}, {Ford}, \&
  {Neill}}]{Ciardullo+1989}
{Ciardullo}, R., {Jacoby}, G.~H., {Ford}, H.~C., \& {Neill}, J.~D. 1989, \apj,
  339, 53

\bibitem[{{Courtois} {et~al.}(2012){Courtois}, {Hoffman}, {Tully}, \&
  {Gottl{\"o}ber}}]{CHTG2012}
{Courtois}, H.~M., {Hoffman}, Y., {Tully}, R.~B., \& {Gottl{\"o}ber}, S. 2012,
  \apj, 744, 43

\bibitem[{{Courtois} {et~al.}(2013){Courtois}, {Pomar{\`e}de}, {Tully},
  {Hoffman}, \& {Courtois}}]{Cosmography}
{Courtois}, H.~M., {Pomar{\`e}de}, D., {Tully}, R.~B., {Hoffman}, Y., \&
  {Courtois}, D. 2013, \aj, 146, 69

\bibitem[{{Courtois} {et~al.}(2009){Courtois}, {Tully}, {Fisher}, {Bonhomme},
  {Zavodny}, \& {Barnes}}]{EDD:HI}
{Courtois}, H.~M., {Tully}, R.~B., {Fisher}, J.~R., {et~al.} 2009, \aj, 138,
  1938

\bibitem[{{Courtois} {et~al.}(2011){Courtois}, {Tully}, \&
  {H{\'e}raudeau}}]{CosmicFlows:HawaiiPhoto}
{Courtois}, H.~M., {Tully}, R.~B., \& {H{\'e}raudeau}, P. 2011, \mnras, 415,
  1935

\bibitem[{{Dalcanton} {et~al.}(2009){Dalcanton}, {Williams}, {Seth}, {Dolphin},
  {Holtzman}, {Rosema}, {Skillman}, {Cole}, {Girardi}, {Gogarten},
  {Karachentsev}, {Olsen}, {Weisz}, {Christensen}, {Freeman}, {Gilbert},
  {Gallart}, {Harris}, {Hodge}, {de Jong}, {Karachentseva}, {Mateo}, {Stetson},
  {Tavarez}, {Zaritsky}, {Governato}, \& {Quinn}}]{Dalcanton+2009}
{Dalcanton}, J.~J., {Williams}, B.~F., {Seth}, A.~C., {et~al.} 2009, \apjs,
  183, 67

\bibitem[{{De Rijcke} {et~al.}(2007){De Rijcke}, {Zeilinger}, {Hau},
  {Prugniel}, \& {Dejonghe}}]{dR+2007}
{De Rijcke}, S., {Zeilinger}, W.~W., {Hau}, G.~K.~T., {Prugniel}, P., \&
  {Dejonghe}, H. 2007, \apj, 659, 1172

\bibitem[{{de Vaucouleurs} {et~al.}(1991){de Vaucouleurs}, {de Vaucouleurs},
  {Corwin}, {Buta}, {Paturel}, \& {Fouqu{\'e}}}]{RC3}
{de Vaucouleurs}, G., {de Vaucouleurs}, A., {Corwin}, Jr., H.~G., {et~al.}
  1991, {Third Reference Catalogue of Bright Galaxies. Volume I: Explanations
  and references. Volume II: Data for galaxies between 0$^{h}$ and 12$^{h}$.
  Volume III: Data for galaxies between 12$^{h}$ and 24$^{h}$.}

\bibitem[{{Di Criscienzo} {et~al.}(2006){Di Criscienzo}, {Caputo}, {Marconi},
  \& {Musella}}]{D+2006}
{Di Criscienzo}, M., {Caputo}, F., {Marconi}, M., \& {Musella}, I. 2006,
  \mnras, 365, 1357

\bibitem[{{Dolphin}(2000)}]{D2000}
{Dolphin}, A.~E. 2000, \apj, 531, 804

\bibitem[{{Feast} \& {Catchpole}(1997)}]{FC1997}
{Feast}, M.~W. \& {Catchpole}, R.~M. 1997, \mnras, 286, L1

\bibitem[{{Fouqu{\'e}} {et~al.}(2003){Fouqu{\'e}}, {Storm}, \&
  {Gieren}}]{Fouque+2003}
{Fouqu{\'e}}, P., {Storm}, J., \& {Gieren}, W. 2003, in Lecture Notes in
  Physics, Berlin Springer Verlag, Vol. 635, Stellar Candles for the
  Extragalactic Distance Scale, ed. D.~{Alloin} \& W.~{Gieren}, 21--44

\bibitem[{{Freedman} {et~al.}(2001){Freedman}, {Madore}, {Gibson}, {Ferrarese},
  {Kelson}, {Sakai}, {Mould}, {Kennicutt}, {Ford}, {Graham}, {Huchra},
  {Hughes}, {Illingworth}, {Macri}, \& {Stetson}}]{HSTKP}
{Freedman}, W.~L., {Madore}, B.~F., {Gibson}, B.~K., {et~al.} 2001, \apj, 553,
  47

\bibitem[{{Gavrilovi{\'c}} {et~al.}(2007){Gavrilovi{\'c}}, {Mickaelian},
  {Petit}, {Popovi{\'c}}, \& {Prugniel}}]{HL:AGN}
{Gavrilovi{\'c}}, N., {Mickaelian}, A., {Petit}, C., {Popovi{\'c}}, L.~{\v C}.,
  \& {Prugniel}, P. 2007, in IAU Symposium, Vol. 238, IAU Symposium, ed.
  V.~{Karas} \& G.~{Matt}, 371--372

\bibitem[{{Gieren} {et~al.}(1998){Gieren}, {Fouqu{\'e}}, \&
  {G{\'o}mez}}]{Gieren+1998}
{Gieren}, W.~P., {Fouqu{\'e}}, P., \& {G{\'o}mez}, M. 1998, \apj, 496, 17

\bibitem[{{Girardi} \& {Salaris}(2001)}]{GS2001}
{Girardi}, L. \& {Salaris}, M. 2001, \mnras, 323, 109

\bibitem[{{Golev} \& {Prugniel}(1998)}]{GP1998}
{Golev}, V. \& {Prugniel}, P. 1998, \aaps, 132, 255

\bibitem[{{Hipparchus}(about 150 BC)}]{Hipparchus}
{Hipparchus}, N. about 150 BC, {Peri megethon kai apostematon}

\bibitem[{{Hubble}(1929)}]{Hubble1929}
{Hubble}, E. 1929, Proceedings of the National Academy of Science, 15, 168

\bibitem[{{Hubble}(1926)}]{Hubble1926}
{Hubble}, E.~P. 1926, \apj, 64, 321

\bibitem[{{Humphreys} {et~al.}(2013){Humphreys}, {Reid}, {Moran}, {Greenhill},
  \& {Argon}}]{HRM2013}
{Humphreys}, E.~M.~L., {Reid}, M.~J., {Moran}, J.~M., {Greenhill}, L.~J., \&
  {Argon}, A.~L. 2013, \apj, 775, 13

\bibitem[{{Ita} \& {Matsunaga}(2011)}]{IM2011}
{Ita}, Y. \& {Matsunaga}, N. 2011, \mnras, 412, 2345

\bibitem[{{Jacobs} {et~al.}(2009){Jacobs}, {Rizzi}, {Tully}, {Shaya},
  {Makarov}, \& {Makarova}}]{EDD:CMD}
{Jacobs}, B.~A., {Rizzi}, L., {Tully}, R.~B., {et~al.} 2009, \aj, 138, 332

\bibitem[{{Jarosik} {et~al.}(2011){Jarosik}, {Bennett}, {Dunkley}, {Gold},
  {Greason}, {Halpern}, {Hill}, {Hinshaw}, {Kogut}, {Komatsu}, {Larson},
  {Limon}, {Meyer}, {Nolta}, {Odegard}, {Page}, {Smith}, {Spergel}, {Tucker},
  {Weiland}, {Wollack}, \& {Wright}}]{WMAP7yr}
{Jarosik}, N., {Bennett}, C.~L., {Dunkley}, J., {et~al.} 2011, \apjs, 192, 14

\bibitem[{{Jensen} {et~al.}(2003){Jensen}, {Tonry}, {Barris}, {Thompson},
  {Liu}, {Rieke}, {Ajhar}, \& {Blakeslee}}]{Jensen+2003}
{Jensen}, J.~B., {Tonry}, J.~L., {Barris}, B.~J., {et~al.} 2003, \apj, 583, 712

\bibitem[{{Jord{\'a}n} {et~al.}(2005){Jord{\'a}n}, {C{\^o}t{\'e}}, {Blakeslee},
  {Ferrarese}, {McLaughlin}, {Mei}, {Peng}, {Tonry}, {Merritt},
  {Milosavljevi{\'c}}, {Sarazin}, {Sivakoff}, \& {West}}]{J+2005}
{Jord{\'a}n}, A., {C{\^o}t{\'e}}, P., {Blakeslee}, J.~P., {et~al.} 2005, \apj,
  634, 1002

\bibitem[{{Kanbur} {et~al.}(2003){Kanbur}, {Ngeow}, {Nikolaev}, {Tanvir}, \&
  {Hendry}}]{Kanbur+2003}
{Kanbur}, S.~M., {Ngeow}, C., {Nikolaev}, S., {Tanvir}, N.~R., \& {Hendry},
  M.~A. 2003, \aap, 411, 361

\bibitem[{{Karachentsev} {et~al.}(2004){Karachentsev}, {Karachentseva},
  {Huchtmeier}, \& {Makarov}}]{CNG}
{Karachentsev}, I.~D., {Karachentseva}, V.~E., {Huchtmeier}, W.~K., \&
  {Makarov}, D.~I. 2004, \aj, 127, 2031

\bibitem[{{Karachentsev} {et~al.}(2009){Karachentsev}, {Kashibadze}, {Makarov},
  \& {Tully}}]{KKMT2009}
{Karachentsev}, I.~D., {Kashibadze}, O.~G., {Makarov}, D.~I., \& {Tully}, R.~B.
  2009, \mnras, 393, 1265

\bibitem[{{Karachentsev} {et~al.}(2013){Karachentsev}, {Makarov}, \&
  {Kaisina}}]{UNGC}
{Karachentsev}, I.~D., {Makarov}, D.~I., \& {Kaisina}, E.~I. 2013, \aj, 145,
  101

\bibitem[{{Kelson} {et~al.}(2000){Kelson}, {Illingworth}, {Tonry}, {Freedman},
  {Kennicutt}, {Mould}, {Graham}, {Huchra}, {Macri}, {Madore}, {Ferrarese},
  {Gibson}, {Sakai}, {Stetson}, {Ajhar}, {Blakeslee}, {Dressler}, {Ford},
  {Hughes}, {Sebo}, \& {Silbermann}}]{K+2000}
{Kelson}, D.~D., {Illingworth}, G.~D., {Tonry}, J.~L., {et~al.} 2000, \apj,
  529, 768

\bibitem[{{Kudritzki} {et~al.}(2008){Kudritzki}, {Urbaneja}, {Bresolin},
  {Przybilla}, {Gieren}, \& {Pietrzy{\'n}ski}}]{FGLR}
{Kudritzki}, R.-P., {Urbaneja}, M.~A., {Bresolin}, F., {et~al.} 2008, \apj,
  681, 269

\bibitem[{{Lanoix} {et~al.}(1999){Lanoix}, {Paturel}, \&
  {Garnier}}]{Lanoix+1999}
{Lanoix}, P., {Paturel}, G., \& {Garnier}, R. 1999, \mnras, 308, 969

\bibitem[{{Lee} {et~al.}(1993){Lee}, {Freedman}, \& {Madore}}]{Lee+1993}
{Lee}, M.~G., {Freedman}, W.~L., \& {Madore}, B.~F. 1993, \apj, 417, 553

\bibitem[{{Madore} \& {Freedman}(1991)}]{Madore+1991}
{Madore}, B.~F. \& {Freedman}, W.~L. 1991, \pasp, 103, 933

\bibitem[{{Makarov} {et~al.}(2006){Makarov}, {Makarova}, {Rizzi}, {Tully},
  {Dolphin}, {Sakai}, \& {Shaya}}]{TRGB1}
{Makarov}, D., {Makarova}, L., {Rizzi}, L., {et~al.} 2006, \aj, 132, 2729

\bibitem[{{Makarov} {et~al.}(2013){Makarov}, {Makarova}, \&
  {Uklein}}]{Makarov+2013}
{Makarov}, D.~I., {Makarova}, L.~N., \& {Uklein}, R.~I. 2013, Astrophysical
  Bulletin, 68, 125

\bibitem[{{McCall}(2014)}]{McCall2014}
{McCall}, M.~L. 2014, \mnras, 440, 405

\bibitem[{{McGaugh} {et~al.}(2000){McGaugh}, {Schombert}, {Bothun}, \& {de
  Blok}}]{M+2000}
{McGaugh}, S.~S., {Schombert}, J.~M., {Bothun}, G.~D., \& {de Blok}, W.~J.~G.
  2000, \apjl, 533, L99

\bibitem[{{Paturel}(1984)}]{P1984}
{Paturel}, G. 1984, \apj, 282, 382

\bibitem[{{Paturel} {et~al.}(1997){Paturel}, {Andernach}, {Bottinelli}, {di
  Nella}, {Durand}, {Garnier}, {Gouguenheim}, {Lanoix}, {Marthinet}, {Petit},
  {Rousseau}, {Theureau}, \& {Vauglin}}]{LEDA7}
{Paturel}, G., {Andernach}, H., {Bottinelli}, L., {et~al.} 1997, \aaps, 124,
  109

\bibitem[{{Paturel} {et~al.}(1988){Paturel}, {Bottinelli}, {Fouque}, \&
  {Gouguenheim}}]{LEDA}
{Paturel}, G., {Bottinelli}, L., {Fouque}, P., \& {Gouguenheim}, L. 1988, in
  European Southern Observatory Conference and Workshop Proceedings, Vol.~28,
  European Southern Observatory Conference and Workshop Proceedings, ed.
  F.~{Murtagh}, A.~{Heck}, \& P.~{Benvenuti}, 435--440

\bibitem[{{Paturel} {et~al.}(1994){Paturel}, {Bottinelli}, \&
  {Gouguenheim}}]{LEDA5}
{Paturel}, G., {Bottinelli}, L., \& {Gouguenheim}, L. 1994, \aap, 286, 768

\bibitem[{{Paturel} {et~al.}(1989){Paturel}, {Fouque}, {Bottinelli}, \&
  {Gouguenheim}}]{PGC}
{Paturel}, G., {Fouque}, P., {Bottinelli}, L., \& {Gouguenheim}, L. 1989,
  \aaps, 80, 299

\bibitem[{{Paturel} {et~al.}(1991){Paturel}, {Garcia}, {Fouque}, \&
  {Buta}}]{LEDA:Size}
{Paturel}, G., {Garcia}, A.~M., {Fouque}, P., \& {Buta}, R. 1991, \aap, 243,
  319

\bibitem[{{Paturel} {et~al.}(2003{\natexlab{a}}){Paturel}, {Petit}, {Prugniel},
  {Theureau}, {Rousseau}, {Brouty}, {Dubois}, \& {Cambr{\'e}sy}}]{HyperLEDA1}
{Paturel}, G., {Petit}, C., {Prugniel}, P., {et~al.} 2003{\natexlab{a}}, \aap,
  412, 45

\bibitem[{{Paturel} {et~al.}(2003{\natexlab{b}}){Paturel}, {Theureau},
  {Bottinelli}, {Gouguenheim}, {Coudreau-Durand}, {Hallet}, \&
  {Petit}}]{HyperLEDA2}
{Paturel}, G., {Theureau}, G., {Bottinelli}, L., {et~al.} 2003{\natexlab{b}},
  \aap, 412, 57

\bibitem[{{Perlmutter} {et~al.}(1999){Perlmutter}, {Aldering}, {Goldhaber},
  {Knop}, {Nugent}, {Castro}, {Deustua}, {Fabbro}, {Goobar}, {Groom}, {Hook},
  {Kim}, {Kim}, {Lee}, {Nunes}, {Pain}, {Pennypacker}, {Quimby}, {Lidman},
  {Ellis}, {Irwin}, {McMahon}, {Ruiz-Lapuente}, {Walton}, {Schaefer}, {Boyle},
  {Filippenko}, {Matheson}, {Fruchter}, {Panagia}, {Newberg}, {Couch}, \& {The
  Supernova Cosmology Project}}]{Perlmutter+99}
{Perlmutter}, S., {Aldering}, G., {Goldhaber}, G., {et~al.} 1999, \apj, 517,
  565

\bibitem[{{Pierce} {et~al.}(2000){Pierce}, {Jurcevic}, \& {Crabtree}}]{PJC2000}
{Pierce}, M.~J., {Jurcevic}, J.~S., \& {Crabtree}, D. 2000, \mnras, 313, 271

\bibitem[{{Prugniel} {et~al.}(1999){Prugniel}, {Golev}, \& {Maubon}}]{PGM1999}
{Prugniel}, P., {Golev}, V., \& {Maubon}, G. 1999, \aap, 346, L25

\bibitem[{{Prugniel} \& {Heraudeau}(1998)}]{PH1998}
{Prugniel}, P. \& {Heraudeau}, P. 1998, \aaps, 128, 299

\bibitem[{{Prugniel} \& {Simien}(1996)}]{PS1996}
{Prugniel}, P. \& {Simien}, F. 1996, \aap, 309, 749

\bibitem[{{Prugniel} {et~al.}(1998){Prugniel}, {Zasov}, {Busarello}, \&
  {Simien}}]{PZBS1998}
{Prugniel}, P., {Zasov}, A., {Busarello}, G., \& {Simien}, F. 1998, \aaps, 127,
  117

\bibitem[{{Radburn-Smith} {et~al.}(2011){Radburn-Smith}, {de Jong}, {Seth},
  {Bailin}, {Bell}, {Brown}, {Bullock}, {Courteau}, {Dalcanton}, {Ferguson},
  {Goudfrooij}, {Holfeltz}, {Holwerda}, {Purcell}, {Sick}, {Streich}, {Vlajic},
  \& {Zucker}}]{Radburn-Smith+2011}
{Radburn-Smith}, D.~J., {de Jong}, R.~S., {Seth}, A.~C., {et~al.} 2011, \apjs,
  195, 18

\bibitem[{{Riess} {et~al.}(1998){Riess}, {Filippenko}, {Challis},
  {Clocchiatti}, {Diercks}, {Garnavich}, {Gilliland}, {Hogan}, {Jha},
  {Kirshner}, {Leibundgut}, {Phillips}, {Reiss}, {Schmidt}, {Schommer},
  {Smith}, {Spyromilio}, {Stubbs}, {Suntzeff}, \& {Tonry}}]{Riess+98}
{Riess}, A.~G., {Filippenko}, A.~V., {Challis}, P., {et~al.} 1998, \aj, 116,
  1009

\bibitem[{{Rizzi} {et~al.}(2007){Rizzi}, {Tully}, {Makarov}, {Makarova},
  {Dolphin}, {Sakai}, \& {Shaya}}]{TRGB2}
{Rizzi}, L., {Tully}, R.~B., {Makarov}, D., {et~al.} 2007, \apj, 661, 815

\bibitem[{{Rozanski} \& {Rowan-Robinson}(1994)}]{RRR1994}
{Rozanski}, R. \& {Rowan-Robinson}, M. 1994, \mnras, 271, 530

\bibitem[{{Saha} {et~al.}(2006){Saha}, {Thim}, {Tammann}, {Reindl}, \&
  {Sandage}}]{Saha+2006}
{Saha}, A., {Thim}, F., {Tammann}, G.~A., {Reindl}, B., \& {Sandage}, A. 2006,
  \apjs, 165, 108

\bibitem[{{Sakai} {et~al.}(1996){Sakai}, {Madore}, \& {Freedman}}]{Sakai+1996}
{Sakai}, S., {Madore}, B.~F., \& {Freedman}, W.~L. 1996, \apj, 461, 713

\bibitem[{{Schlegel} {et~al.}(1998){Schlegel}, {Finkbeiner}, \&
  {Davis}}]{Schlegel+1998}
{Schlegel}, D.~J., {Finkbeiner}, D.~P., \& {Davis}, M. 1998, \apj, 500, 525

\bibitem[{{Storm} {et~al.}(2011){Storm}, {Gieren}, {Fouqu{\'e}}, {Barnes},
  {Soszy{\'n}ski}, {Pietrzy{\'n}ski}, {Nardetto}, \& {Queloz}}]{S+2011}
{Storm}, J., {Gieren}, W., {Fouqu{\'e}}, P., {et~al.} 2011, \aap, 534, A95

\bibitem[{{Tammann} {et~al.}(2003){Tammann}, {Sandage}, \&
  {Reindl}}]{Tammann+2003}
{Tammann}, G.~A., {Sandage}, A., \& {Reindl}, B. 2003, \aap, 404, 423

\bibitem[{{Tonry} \& {Schneider}(1988)}]{SBF}
{Tonry}, J. \& {Schneider}, D.~P. 1988, \aj, 96, 807

\bibitem[{{Tonry} {et~al.}(2000){Tonry}, {Blakeslee}, {Ajhar}, \&
  {Dressler}}]{Tonry+2000}
{Tonry}, J.~L., {Blakeslee}, J.~P., {Ajhar}, E.~A., \& {Dressler}, A. 2000,
  \apj, 530, 625

\bibitem[{{Toomer}(1974)}]{Toomer1974}
{Toomer}, G. 1974, Archives for the History of the Exact Sciences, 14, 126

\bibitem[{{Tully} \& {Courtois}(2012)}]{TC2012}
{Tully}, R.~B. \& {Courtois}, H.~M. 2012, \apj, 749, 78

\bibitem[{{Tully} {et~al.}(2013){Tully}, {Courtois}, {Dolphin}, {Fisher},
  {H{\'e}raudeau}, {Jacobs}, {Karachentsev}, {Makarov}, {Makarova},
  {Mitronova}, {Rizzi}, {Shaya}, {Sorce}, \& {Wu}}]{CosmicFlows2}
{Tully}, R.~B., {Courtois}, H.~M., {Dolphin}, A.~E., {et~al.} 2013, \aj, 146,
  86

\bibitem[{{Tully} \& {Fisher}(1977)}]{TullyFisherRelation}
{Tully}, R.~B. \& {Fisher}, J.~R. 1977, \aap, 54, 661

\bibitem[{{Tully} \& {Pierce}(2000)}]{TP2000}
{Tully}, R.~B. \& {Pierce}, M.~J. 2000, \apj, 533, 744

\bibitem[{{Tully} {et~al.}(2009){Tully}, {Rizzi}, {Shaya}, {Courtois},
  {Makarov}, \& {Jacobs}}]{EDD}
{Tully}, R.~B., {Rizzi}, L., {Shaya}, E.~J., {et~al.} 2009, \aj, 138, 323

\bibitem[{{Tully} {et~al.}(2008){Tully}, {Shaya}, {Karachentsev}, {Courtois},
  {Kocevski}, {Rizzi}, \& {Peel}}]{TSK2008}
{Tully}, R.~B., {Shaya}, E.~J., {Karachentsev}, I.~D., {et~al.} 2008, \apj,
  676, 184

\bibitem[{{Udalski} {et~al.}(1999){Udalski}, {Szymanski}, {Kubiak},
  {Pietrzynski}, {Soszynski}, {Wozniak}, \& {Zebrun}}]{Udalski+1999}
{Udalski}, A., {Szymanski}, M., {Kubiak}, M., {et~al.} 1999, \actaa, 49, 201

\bibitem[{{V{\'e}ron-Cetty} \& {V{\'e}ron}(2010)}]{VeronCat13}
{V{\'e}ron-Cetty}, M.-P. \& {V{\'e}ron}, P. 2010, \aap, 518, A10

\end{thebibliography}

\label{lastpage}

\end{document}